\documentclass[twocolumn]{aastex631}
\usepackage{color}
\usepackage{xspace}
\usepackage[caption=false]{subfig}
\usepackage{amsmath}

\newcommand{\nic}{{\it NICER}\xspace}
\newcommand{\nus}{{\it NuSTAR}\xspace}
\newcommand{\suzaku}{{\it Suzaku}\xspace}

\newcommand{\chandra}{{\it Chandra}\xspace}

\newcommand{\ls}{LS~5039\xspace}

\newcommand{\psrb}{PSR~B1259-63\xspace}

\begin{document}

\title{Unveiling properties of the non-thermal X-ray production in the gamma-ray binary LS~5039\\
using the long-term pattern of its fast X-ray variability}

\correspondingauthor{Hiroki Yoneda}
\email{hiroki.yoneda@riken.jp}

\author[0000-0002-5345-5485]{Hiroki Yoneda}
\affiliation{RIKEN Nishina Center, 2-1 Hirosawa, Wako, Saitama 351-0198, Japan}
\affiliation{Julius-Maximilians-Universit\"{a}t W\"{u}rzburg, Fakult\"{a}t f\"{u}r Physik und Astronomie, Institut f\"{u}r Theoretische Physik und Astrophysik, Lehrstuhl f\"{u}r Astronomie, Emil-Fischer-Str. 31, D-97074 W\"{u}rzburg, Germany}

\author[0000-0002-6043-5079]{Valenti Bosch-Ramon}
\affiliation{Departament de Física Quàntica i Astrofísica, Institut de Ciències del Cosmos (ICC), Universitat de Barcelona (IEEC-UB), Martí i Franquès 1, E08028 Barcelona, Spain}

\author[0000-0003-1244-3100]{Teruaki Enoto}
\affiliation{Extreme natural phenomena RIKEN Hakubi Research Team, Cluster for Pioneering Research, RIKEN, Hirosawa 2-1, Wako, Saitama, 351-0198, Japan}

\author[0000-0002-7576-7869]{Dmitry Khangulyan}
\affiliation{Graduate School of Artificial Intelligence and Science, Rikkyo University, Nishi-Ikebukuro 3-34-1, Toshima-ku, Tokyo 171-8501, Japan}

\author[0000-0002-5297-5278]{Paul S. Ray}
\affiliation{U.S. Naval Research Laboratory, Washington, DC 20375, USA}

\author[0000-0001-7681-5845]{Tod Strohmayer}
\affiliation{Astrophysics Science Division and Joint Space-Science Institute, NASA Goddard Space Flight Center, Greenbelt, MD 20771, USA}

\author{Toru Tamagawa}
\affiliation{RIKEN Nishina Center, 2-1 Hirosawa, Wako, Saitama 351-0198, Japan}

\author[0000-0002-9249-0515]{Zorawar Wadiasingh}
\affiliation{Astrophysics Science Division, NASA/GSFC, Greenbelt, MD 20771, USA}
\affiliation{Department of Astronomy, University of Maryland, College Park, Maryland 20742, USA}
\affiliation{Center for Research and Exploration in Space Science and Technology, NASA/GSFC, Greenbelt, Maryland 20771, USA}

\begin{abstract}
Gamma-ray binary systems, a subclass of high-mass X-ray binaries, show non-thermal emissions from radio to TeV. While efficient electron acceleration is considered to take place in them, the nature of the acceleration mechanism and the physical environments in these systems have been a long-standing question.
In this work, we report on long-term recurrent patterns in the short-term variability of the soft X-ray emission of \ls, one of the brightest gamma-ray binary systems.
The Neutron star Interior Composition Explorer (\nic) observed \ls four times from 2018 to 2021.
By comparing them with the previous \suzaku and \nus long-exposure observations, we studied the long-term evolution of the orbital light curve in the soft X-ray band.
Although the observations by \nic and \suzaku are separated by $\sim$14 years, i.e., more than \(10^3\) orbits, the orbital light curves show remarkable consistency after calculating their running averages with a window width $\gtrsim 70$~ks.
Furthermore, all of the light curves show short-term variability with a time scale of $\sim$10 ks.
Since the column density did not vary when the flux changed abruptly, such a short-term variability seems to be an intrinsic feature of the X-ray emission.
We propose that the short-term variability is caused by clumps (or inhomogeneities) of the companion star wind impacting the X-ray production site.
The observed time scale matches well with the lifetime of the clumps interacting with the pulsar wind and the dynamical time scale of the relativistic intrabinary shock in the pulsar wind scenario. 
\end{abstract}


\section{Introduction} \label{sec_intro}
Classical gamma-ray binary systems are a subclass of high-mass X-ray binaries characterized by their strong non-thermal emission from the radio to the TeV band \citep[e.g.,][]{Dubus2013,Paredes2019}.
Following highly sensitive observations in the GeV and TeV bands, this new class of binary systems was established in the middle of the 2000s \citep{hessls50392005,PSRB1259HESS2005,2006Sci...312.1771A,2008ApJ...679.1427A}.
Their spectra in the X-ray band are well described with a hard power-law function, and typically they peak beyond 1 MeV. These features could be a sign of extreme electron acceleration in the presence of a dense and hot target photon field \citep[see, e.g.,][]{khangulyan2008,takahashistudy2009}.

The number of sources identified as classical gamma-ray binaries so far is approximately ten \citep[see, e.g.,][and references therein]{Paredes2019}. They consist of an OB star and a compact object, and, except for a few cases, the nature of the compact object remains unknown.
Thus, several scenarios have been proposed, with different assumptions for the nature of the compact object, to interpret the origin of the non-thermal emission in these sources.
A popular scenario is that of a non-accreting pulsar with a strong relativistic wind powering the non-thermal emission in these systems \citep[e.g.,][]{Dubus2006b,Zabalza2013,takata2014,dubus2015}.
In this scenario, the relativistic pulsar wind and the strong stellar wind collide, forming a shocked two-wind region where electrons can be accelerated \citep[see, e.g.,][]{Bogovalov2008} and produce non-thermal emission via the synchrotron and inverse Compton processes \citep{1997ApJ...477..439T,1999APh....10...31K,2007MNRAS.380..320K}.
The microquasar scenario has also been studied as a possible alternative, in which gamma rays are produced in a relativistic jet launched from a stellar-mass black hole \citep{Paredes2000,Paredes2006,khangulyan2008}.
Another potential scenario has been proposed in which some gamma-ray binaries may contain magnetars, proposal based on a magnetar-like X-ray burst \citep{torresmagnetarlike2012}, and on hints of 9-second pulsations in the hard X-ray band \citep{yoneda2020}. 

As one of the brightest gamma-ray binaries in the Galaxy,
\ls has been studied intensively.
The system has a $\approx 3.9$~day period, its companion star is a main sequence O star with a mass of $23 M_\odot$ \citep{casarespossible2005}, and it is still unknown whether the compact object is a neutron star or a black hole.
Its broad-band spectrum shows strong synchrotron emission that seems to peak around a few tens of MeV \citep{collmarls2014,Falanga2021,Yoneda_2021}, which suggests that particle acceleration operating in \ls is extremely efficient, close to the limit allowed by electrodynamics 
\citep{khangulyan2008,takahashistudy2009}.

A remarkable feature of \ls is the periodic variability of the X-ray emission over the years \citep{BoschRamon2005,takahashistudy2009,kishishitalongterm2009}.
Particularly, \citet{kishishitalongterm2009} analyzed the X-ray data obtained from 1999 to 2007 using different satellites ({\it ASCA}, {\it XMM-Newton}, {\it Chandra}, and \suzaku) and found that the orbital modulation in that period seems to be very stable, almost clock-like, a finding reinforced by evidence of orbital modulation also found in {\it RXTE} data \citep{BoschRamon2005}.
Also, fine structures, such as spikes in the orbital light curve, possibly repeat orbit-to-orbit. These results may support the pulsar wind scenario if they can be associated to the pulsar-stellar wind interaction. However, except for a few cases, most of the mentioned observations covered short orbital phase intervals. Thus, orbital-period-long observations 
are essential to confirm the presence of a long-term recurrence of fast variability, which may unveil information of the two-wind interaction structure. In this regard, data in the soft X-ray band are distinctly important as it can provide with the highest statistics and thus the richest information.

In this paper,
we analyze \ls observations in the soft X-ray band performed in four runs by \nic
and report the properties of the X-ray synchrotron emission from \ls for all the orbital phases, as \nic observations cover the whole orbit.
These data allow us to study both the short- and long-term variability of the soft X-ray emission by comparison with previous observations.
In Section~\ref{sec_observation}, we describe the four observations of \nic and the data reduction.
Then, Section~\ref{sec_orbitallightcurve} shows the results of the orbital light curves obtained by \nic and \suzaku.
As a reference, we also show the orbital light curve in hard X-rays using the \nus observation from 2016.
In Section~\ref{sec_discussion}, we discuss the short-term variability of the orbital light curve in the context of the interaction between a pulsar and a stellar wind, a scenario that may explain well the observed behavior, and
comment on the 9-second pulsation reported by \cite{yoneda2020} and whose detection is still under debate \citep{Volkov2021}. 
Finally, in Section~\ref{sec_conclusion}, the conclusions of this work are presented.

\section{Observations and data reduction} \label{sec_observation}

\cite{kishishitalongterm2009} reported that the orbital modulation in soft X-rays ($\lesssim 10$~keV) seen in LS~5039 showed a remarkable stability, which was revealed by comparing X-ray observations taken several years apart.
In their work, however, except for the \suzaku observation, each observation covered a small orbital phase interval, and the discussion of the long-term variability was limited to orbital phases close to the apastron and inferior conjunctions. More recent hard X-ray observations by \nus covering all orbital phases were also carried out \citep{Yoneda_2021,Volkov2021}, but the somewhat different energy range made a direct comparison with previous X-ray data less straightforward (see below). 
In this context, stimulated by the prospect of confirming previous results, we observed \ls with \nic four times from 2018 to 2021, and the combination of these observations provides us with soft X-ray data covering the entire orbit of \ls.

\nic is an International Space Station (ISS) payload for X-ray observations in the range 0.2--12 keV \citep{Gendreau2016}.  
X-rays are measured by \nic's main science instrument, the X-ray Timing Instrument (XTI).
It consists of 56 pairs of X-ray optics and a silicon drift detector.
With this design, \nic achieves a large effective area of 1900~cm$^2$ at 1.5 keV with a timing resolution finer than 300~ns. The average count rate of the signal from \ls was about one count per second. The relevant information on \nic observations of \ls is given in Table~\ref{tab_obs_info}.

The data reduction and analysis were performed with the {\tt NICERDAS} version 2022-01-17\_V009 and \nic~{\tt CALDB} version xti20210707.
The spectral fitting was performed with XSPEC version 12.12.1 \citep{xspec}.
During the observation started on August 26, 2019 (OBSID: 4631010201--2), \nic was close to the South Atlantic Anomaly (SAA), and the first and last parts of the observation suffered from higher background rates than usual.
Thus, we excluded these time intervals to avoid systematic uncertainty due to large background events (MET 241435000--241462000 and MET 241517000--241556000).
We note that \nic also observed \ls on March 12, 2019 (OBSID: 2030170101), but its net exposure was not long enough ($<$ 2~ks) to be used in this work.
After the data reduction, the data set of the \nic observations has a net and gross exposure of 105.9~ks and 444.5~ks, respectively (see Table~\ref{tab_obs_info} for the exposure of each observation).
In this work, the background rate and spectra were estimated by the 3C50 model, an empirical background model by \cite{3c50_2022}.
The response and auxiliary response files were produced using the tools {\tt nicerrmf} and {\tt nicerarf}, respectively.

\section{Spectral analysis} \label{sec_orbitallightcurve}

\subsection{Phase-resolved spectral analysis of the \nic data} \label{subsec_spectrumanalysis}

In the spectral analysis, we modeled the spectra as a single power-law function with a photoelectric absorption model ({\tt phabs} in XSPEC).
We set the energy range to 1.5--5.0 keV in the spectral fitting.
We note that below $\sim$1~keV, the background dominates over the source events. To avoid systematic uncertainties due to the background modeling, we set the lower limit of the fitting energy range as 1.5~keV.

Table~\ref{tab_obs_info} describes the results of the spectral fitting for the four observations.
The column density ranges from $0.67\times10^{21}$ cm$^{-2}$ to $0.80\times10^{21}$ cm$^{-2}$.
Since their differences are within their statistical errors (see Table~\ref{tab_obs_info}), the column density does not show significant variations along the orbit. 
The obtained values of the column density and the photon index are also consistent with the previous observations \citep{Bosch-Ramon2007,takahashistudy2009,kishishitalongterm2009}.

For the first three observations, we divided the observations into subsets, each covering $\sim0.15$ orbital phase intervals, to investigate the spectral change on a shorter time scale.
Again, their comparison does not reveal any significant change in the column density.
For the photon index, especially the observation in March 2019, we can see the trend that it gets the smallest around apastron, that is, at orbital phase of $0.5$. This behavior is also consistent with the results of \cite{BoschRamon2005,takahashistudy2009,kishishitalongterm2009}. 

As the default choice, we adopted the orbital parameters reported by \cite{casarespossible2005} to allow a direct comparison with the results reported in \cite{kishishitalongterm2009}. We also performed cross checks adopting the slightly different orbital parameters reported in \cite{Aragona2010,Sarty2011}. We discuss
the uncertainty due to different orbital solutions in Appendix \ref{sec_unc_orbparam}. 

\begin{table*}[!htbp]
  \caption{List of the NICER observations and the spectral fitting result in each orbital phase interval} 
  \label{tab_obs_info}
  \begin{center}
  \begin{tabular}{cccccccc} \hline
  Date & Observation ID & $\phi^{\mathrm{a}}$ & Exposure (ks)$^{\mathrm{b}}$ & $N_{\mathrm{H}}$$^{\mathrm{c}}$ & $\Gamma$$^{\mathrm{d}}$ & Flux (1.5--5 keV)$^{\mathrm{e}}$ & $\chi^2$/dof \\ \hline
  2018 Oct 11 & 1030170101--2 & 0.88--1.16 & 16.6/96.0 & $0.75 \pm 0.14$ & $1.57 \pm 0.10$ & $3.38\pm 0.11$ & $230.1/198$\\
& & 0.88--1.02 &  
& $0.63 \pm 0.18$
& $1.45 \pm 0.13$
& $4.06 \pm 0.17$
& $142.7/150$ \\
& & 1.02--1.16 &  
& $0.82 \pm 0.27$
& $1.66 \pm 0.21$
& $2.77 \pm 0.19$
& $49.6/69$ \\
  2019 Mar 13 & 2030170102--4 & 0.97--1.57 & 41.3/199.5 & $0.80 \pm 0.10$ & $1.74\pm 0.08$ & $3.21 \pm 0.08$ & $371.0/313$ \\
& & 0.97--1.14 &  
& $1.11 \pm 0.20$
& $2.12 \pm 0.16$
& $2.66 \pm 0.15$
& $196.3/172$ \\
& & 1.14--1.29 &  
& $0.77 \pm 0.21$
& $1.83 \pm 0.16$
& $2.62 \pm 0.14$
& $126.2/146$ \\
& & 1.29--1.44 &
& $0.62 \pm 0.15$
& $1.52 \pm 0.11$
& $3.76 \pm 0.13$
& $191.1/176$ \\
& & 1.44--1.57 &  
& $0.71 \pm 0.23$
& $1.45 \pm 0.17$
& $4.72 \pm 0.26$
& $73.8/90$ \\
  2021 Aug 22 & 4631010101--2 & 0.61--0.90 & 32.2/97.1 & $0.68 \pm 0.08 $ & $1.62 \pm 0.06$ & $5.16 \pm 0.10$ & $367.9/321$ \\
& & 0.61--0.75 &  
& $0.75 \pm 0.12$
& $1.64 \pm 0.09$
& $5.10 \pm 0.15$
& $293.5/238$ \\
& & 0.75--0.90 &  
& $0.62 \pm 0.10$
& $1.62 \pm 0.08$
& $5.13 \pm 0.13$
& $266.2/253$ \\
  2021 Aug 26 & 4631010201--2 & 0.70--0.85 & 15.8/51.9$^{\mathrm{f}}$ & $0.67 \pm 0.11$ & $1.60 \pm 0.09$ & $4.82 \pm 0.14$ & $224.1/239$ \\ 
\hline
  \multicolumn{2}{c}{all} & & 105.9/444.5 & $0.74 \pm 0.05$ & $1.65 \pm 0.04$ & $4.11 \pm 0.05$ & $448.0/349$\\ \hline
  \end{tabular}
  \end{center}
  {Notes. Errors correspond to 1 $\sigma$ confidence interval. \\
$^{\mathrm{a}}$ Orbital phase intervals analyzed.\\
$^{\mathrm{b}}$ The former and latter values correspond to the net and gross exposures, respectively.\\
$^{\mathrm{c}}$ The column density with an unit of $10^{22}~\mathrm{cm^{-2}}$.\\
$^{\mathrm{d}}$ The photon index of the power-law function.\\
$^{\mathrm{e}}$ The unabsorbed flux with an unit of $10^{-12}~\mathrm{erg~cm^{-2}~s^{-1}}$.\\
$^{\mathrm{f}}$ The time intervals of MET241435000--241462000 and MET241517000--241556000 were not used in the analysis, and they were not included in the net and gross exposures shown here.}
\end{table*}

\subsection{Comparison between the 2007 \suzaku and 2018--2021 \nic observations}\label{subsec_suzaku}

Here we investigate the behavior of the soft X-ray emission over the whole orbit using \nic and \suzaku observations. To carry out the comparison, each \nic observation was divided into subsets of 4~ks exposure intervals, which approximately corresponds to a 1\%  of the orbit, to calculate the 1.5--5.0~keV flux.
We again fitted the spectrum in each interval using a single power-law function with a photoelectric absorption model. Since the number of events in each interval is small, the C-statistic was adopted for the spectral fitting. We also fixed the column density to $0.74 \times 10^{21}$ cm$^{-2}$, which is derived from all of the \nic observations (see the bottom of Table~\ref{tab_obs_info}). We note that the column density has a typical uncertainty of $\sim 1\times10^{20}$ cm$^{-2}$, but the translated systematic error in the flux is about half of its statistical error. Thus, the obtained results are not affected by fixing the column density. 
In order to remove data points with a large statistical/systematic error,
we excluded time intervals that satisfy any of the following two conditions. One is that the net exposure of the subset is less than 400~s, i.e., $<10$\% of the gross one. The other is that the background rate in the range 1.5--5.0~keV is larger than 0.45~Hz, which is about twice as large as the average background rate.

To compare the \nic results with the other available soft X-ray observation covering a whole orbit, we also analyzed the \suzaku/XIS data using the events in the same energy range (1.5--5.0~keV). The data reduction and analysis methods followed are those used in \cite{takahashistudy2009, Yoneda_2021}.
We note that \suzaku observations of \ls started from the orbital phase ($\phi$) of 0.0 and covered approximately 1.5 orbits.
In the following figures, we present the analyzed data by separating the first orbit and the last half.

\begin{figure*}
\centering
\includegraphics[width = 0.70\textwidth]{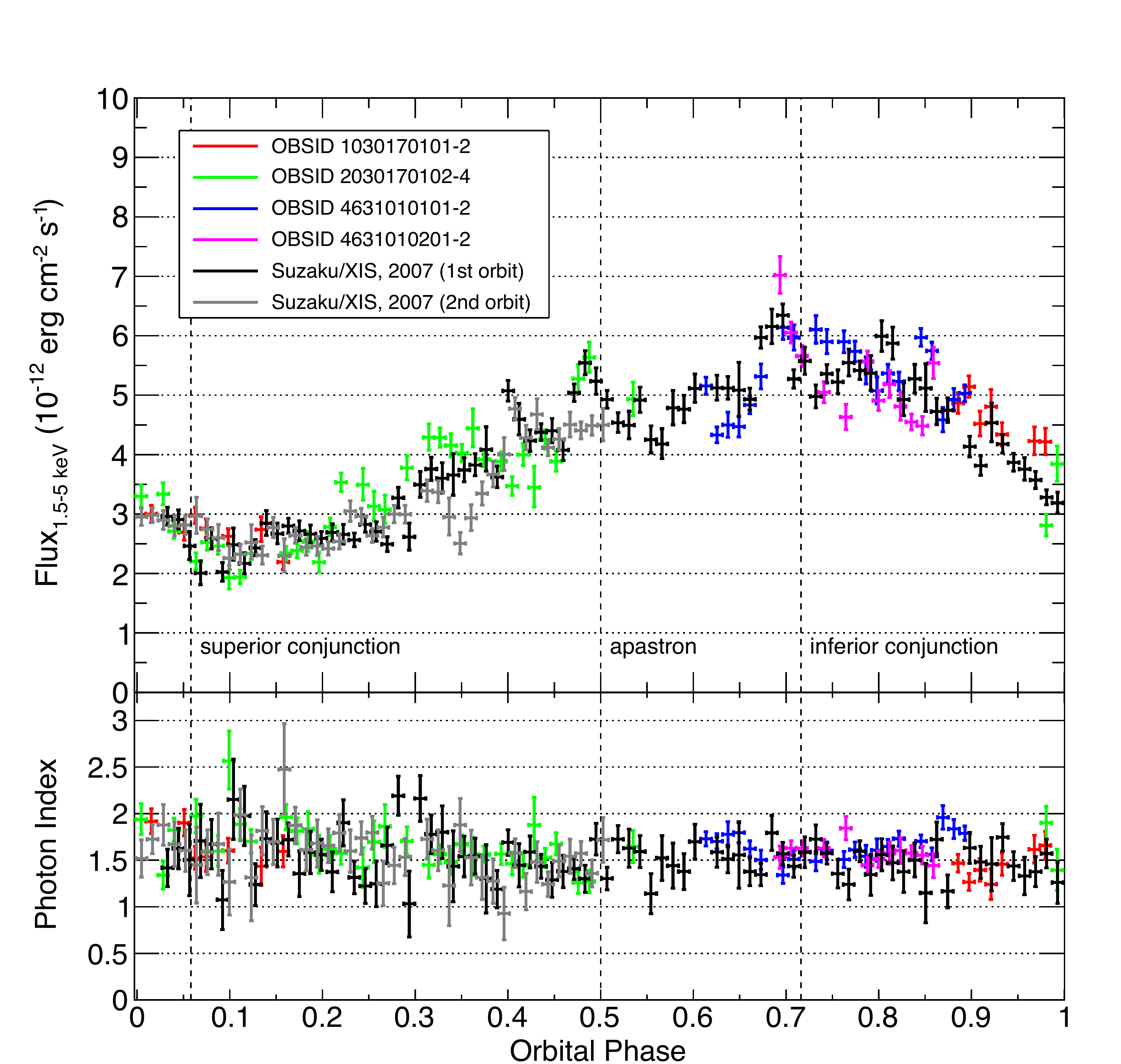}
\caption{Orbital light curves for the flux (top panel) and the photon index (bottom panel) obtained from \nic (four observations) and \suzaku data. The bin width is 4~ks. The data points of \suzaku observations colored in black correspond to the first orbit. The energy range in the fitting is set to 1.5--5.0~keV. Vertical dashed lines indicate inferior and superior conjunctions of the compact object, and apastron.}
\label{fig_orbital_light_curve}
\end{figure*}

Figure~\ref{fig_orbital_light_curve} shows the obtained orbital light curves of \nic and \suzaku with a bin width of 4 ks.
The top and bottom panels show the flux and the photon index in the range 1.5--5.0 keV, respectively.
A different color distinguishes each \nic observation, and the black- and grey-colored data points correspond to the consequent 1.5 orbits observed by \suzaku/XIS in 2007.
The orbital light curves of the two instruments show a similar trend: the X-ray flux is at its lowest at $\phi \simeq 0.1$ around the superior conjunction of the compact object, and at its highest at $\phi \simeq 0.7$ around the inferior conjunction \citep{casarespossible2005}.
Due to the relatively large statistical errors, we do not see any sudden significant change in the photon index, and its value lies mostly between 1 and 2.
Interestingly, the fluxes obtained by \nic and \suzaku are consistent within $\sim 1\times 10^{-12}~\mathrm{erg~cm^{-2}~s^{-1}}$. Moreover, the light curves show a short-term variability with a time scale of a few tens of ks, for instance, at $\phi \sim 0.48$ in the \suzaku data, at $\phi \sim 0.23$ in the \nic data (green), etc.
These results imply that the orbital modulation in the soft X-ray band is stable over a time scale of $\sim$14 years, while on the other hand there are 
sudden flux changes on scales of $\sim$10 ks. The short-term variability appears to be similar both when comparing the two consecutive orbits observed by \suzaku and two orbits separated by $\sim14$ years. To examine the long-term stability of the orbital modulation, in the rest of this subsection we compare the running-averaged orbital light curves. Then, in the following subsection, we focus on the short-term variability.

\begin{figure}
\centering
\includegraphics[width = 0.45\textwidth]{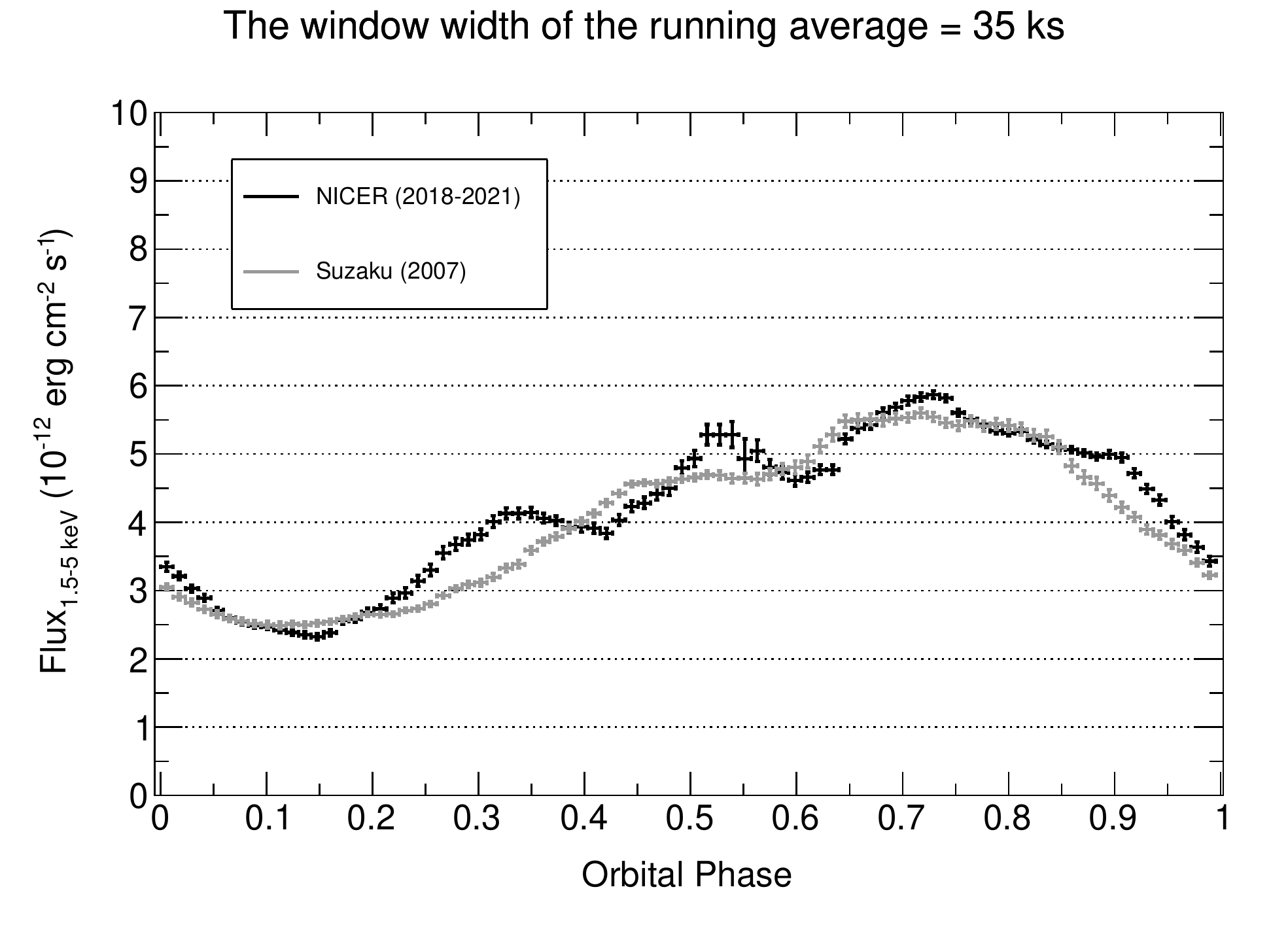}
\includegraphics[width = 0.45\textwidth]{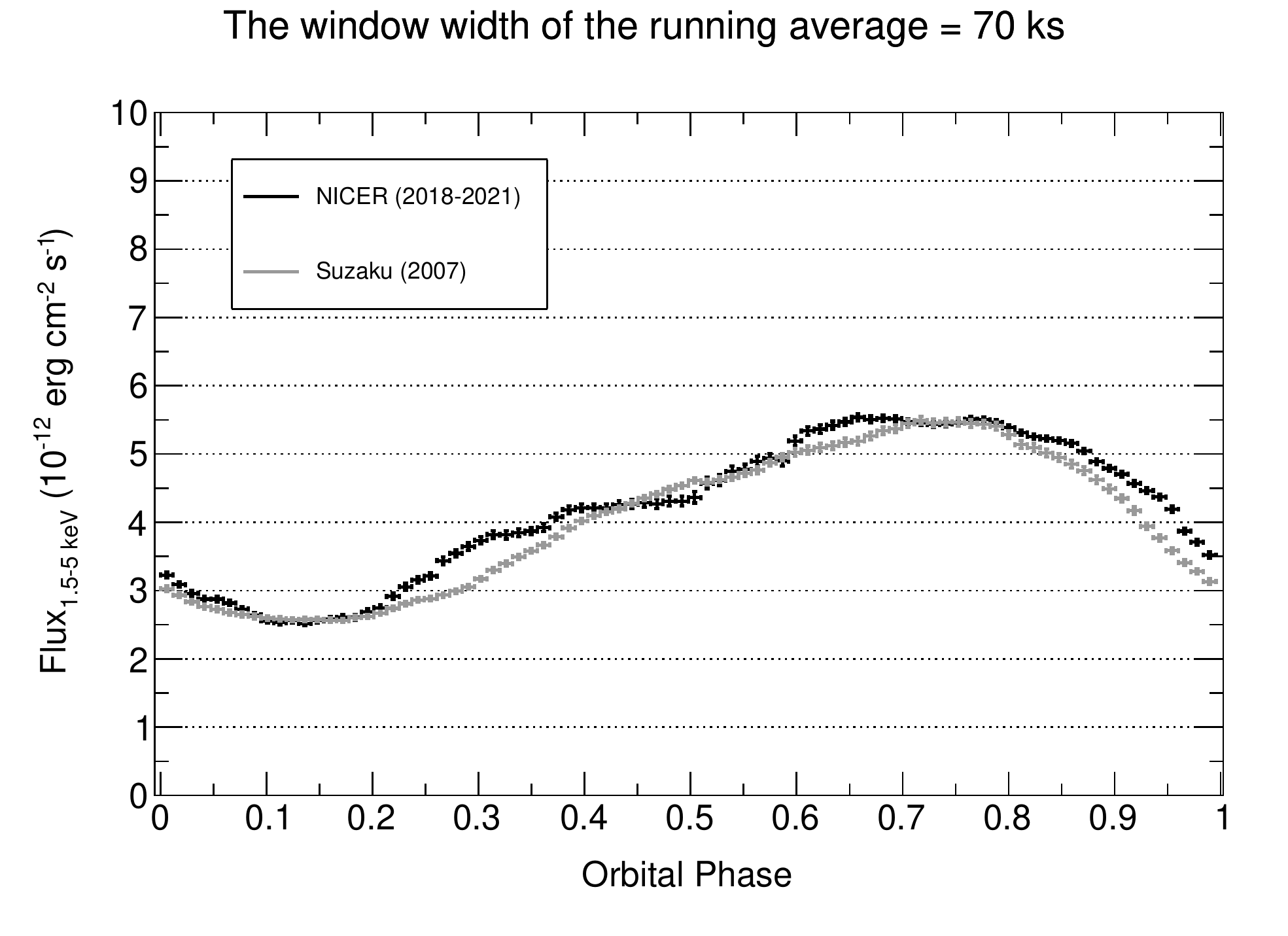}
\includegraphics[width = 0.45\textwidth]{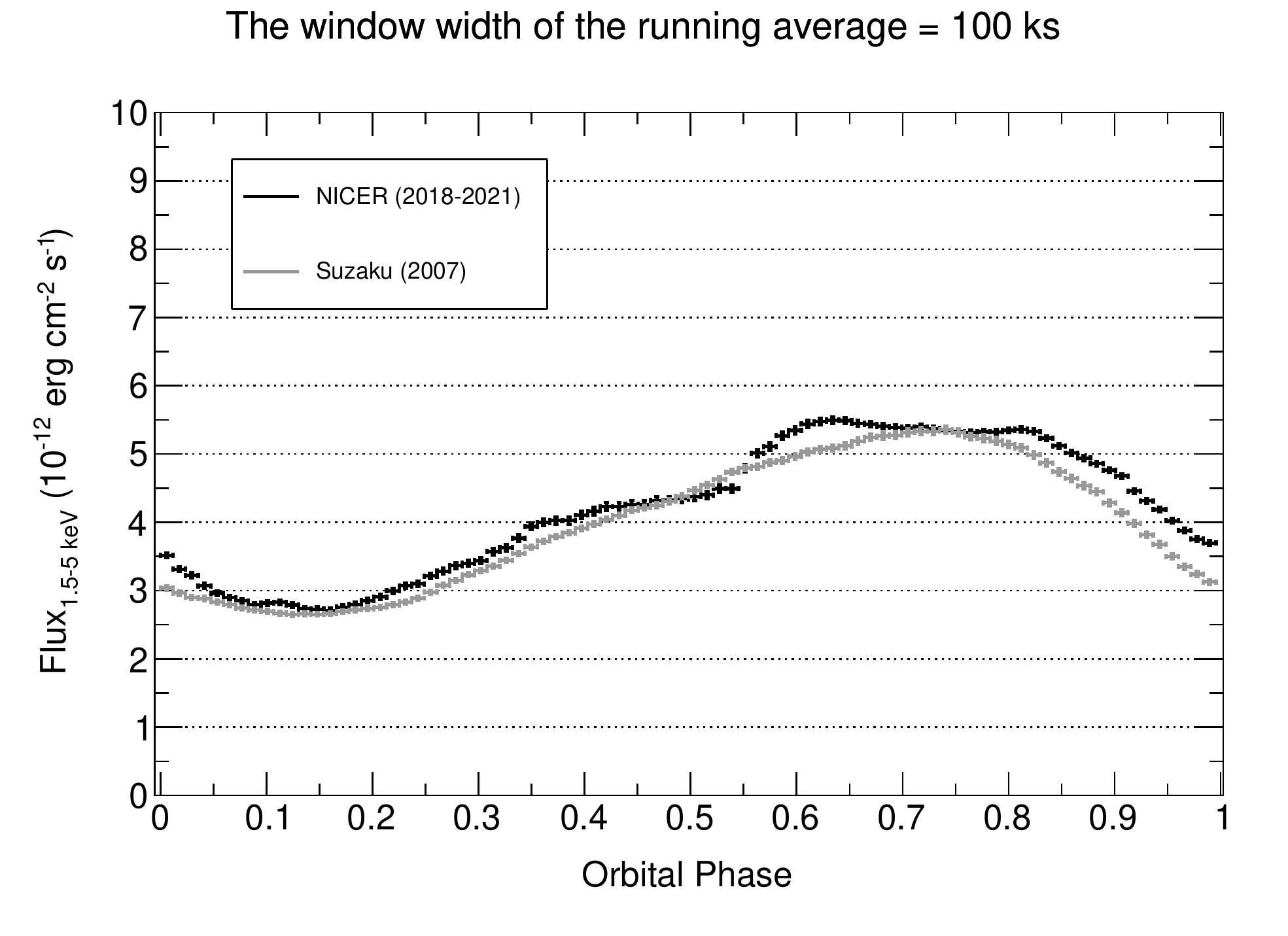}
\caption{Running-averaged orbital light curves of the \suzaku and \nic observations. Different window widths of the running average are applied to them: 35~ks (top), 70~ks (middle), and 100~ks (bottom).}
\label{fig_orbital_light_averaged}
\end{figure}

Figure~\ref{fig_orbital_light_averaged} shows the running-averaged orbital light curves of \nic and \suzaku observations. The running-averaged flux was calculated extracting flux data points within a certain window width for orbital phases separated by 4~ks and calculated their average; the adopted window widths were: 35, 70, and 100 ks, which correspond to $\phi$ intervals $\sim 0.1, 0.2$, and $0.3$, respectively.
While the \nic and \suzaku running-averaged orbital light curves show some differences for a window width of 35 ks, they show a remarkable agreement for a window width of 70 ks or longer.
In particular, at $\phi \sim 0.1-0.2$ and $0.7-0.8$ the two curves are almost identical. In these orbital phase intervals, the maximum flux difference is only $\approx 3$\% for a window width of 70 ks, and even if we compare the fluxes in all phases the maximum difference is only $\approx 18$\% (at $\phi \sim 0.3$). We note that at $\phi \sim 0.3$, the \nic observation was performed only once in 2019, and the light curve without applying the running average already showed a relatively large variability. We note that, despite no robust conclusion can be drawn from the following fact since it may be just a coincidence, when adopting the orbital parameters from \cite{aragonaorbits2009} the agreement between the two light curves is even more apparent around $\phi = 0.7 - 1.1$, as shown in Figure~\ref{fig_orbital_light_averaged_aragona} (see Appendix~\ref{sec_unc_orbparam}). 

\begin{figure}
\centering
\includegraphics[width = 0.45\textwidth]{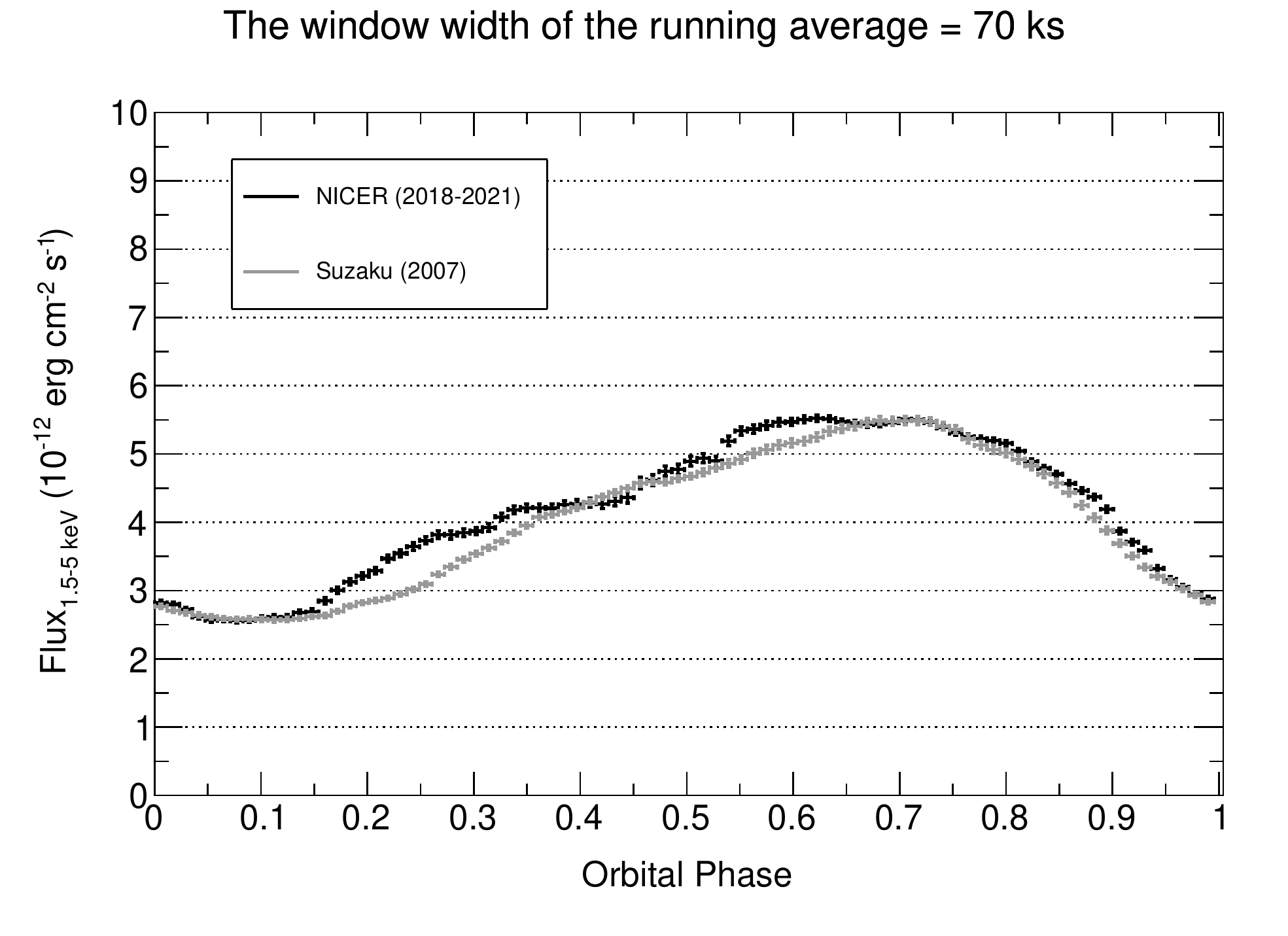}
\caption{The same as in the middle panel of Figure~\ref{fig_orbital_light_averaged}, but for the orbital solution from \cite{aragonaorbits2009}.}
\label{fig_orbital_light_averaged_aragona}
\end{figure}

\subsection{Short-term variability with a time scale of a few tens of ks}

\begin{figure*}
\gridline{\fig{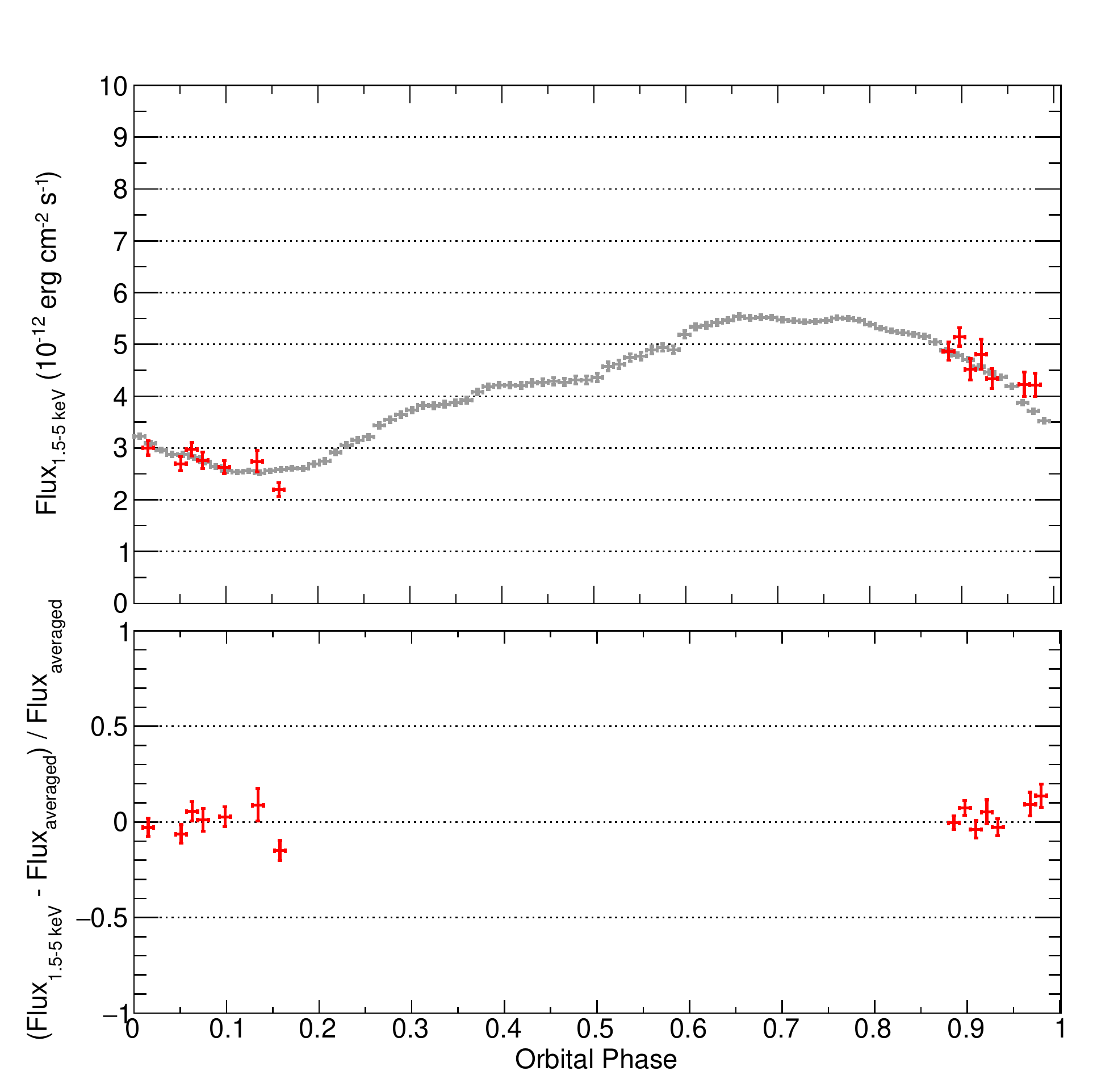}{0.4\textwidth}{(a) OBSID 1030170101--2}
          \fig{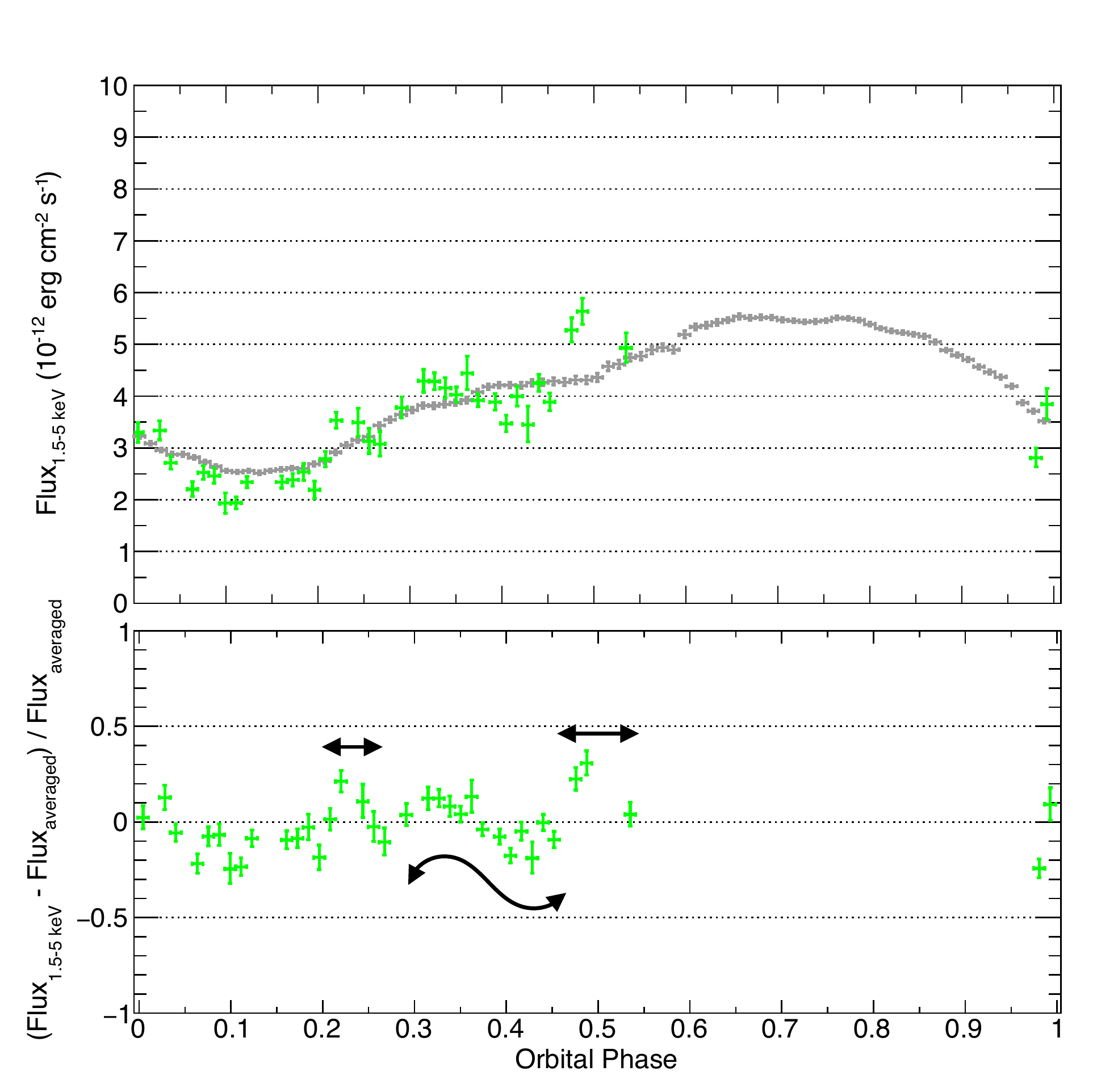}{0.4\textwidth}{(b) OBSID 2030170102--4}}
\gridline{\fig{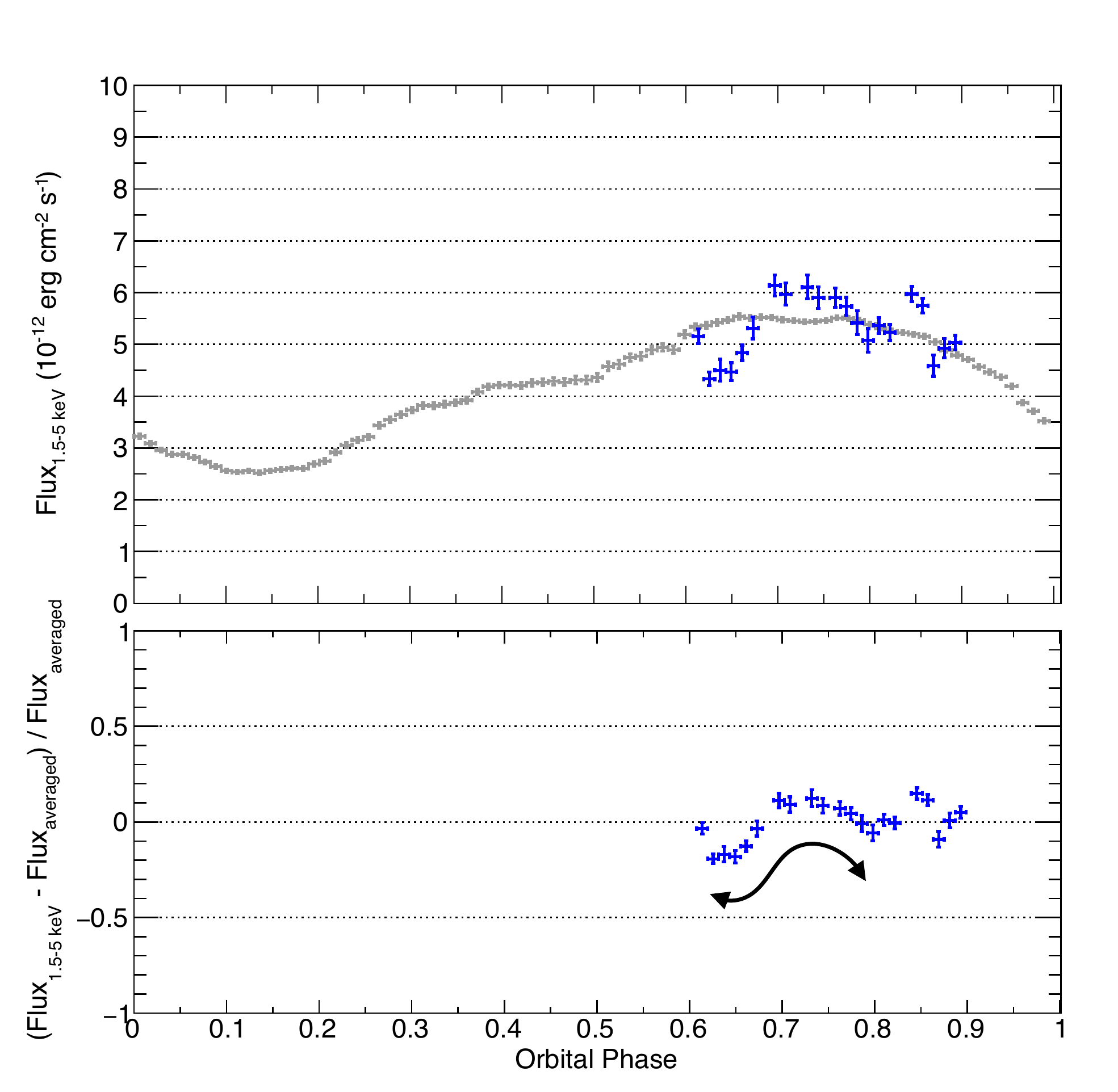}{0.4\textwidth}{(c) OBSID 4631010101--2}
          \fig{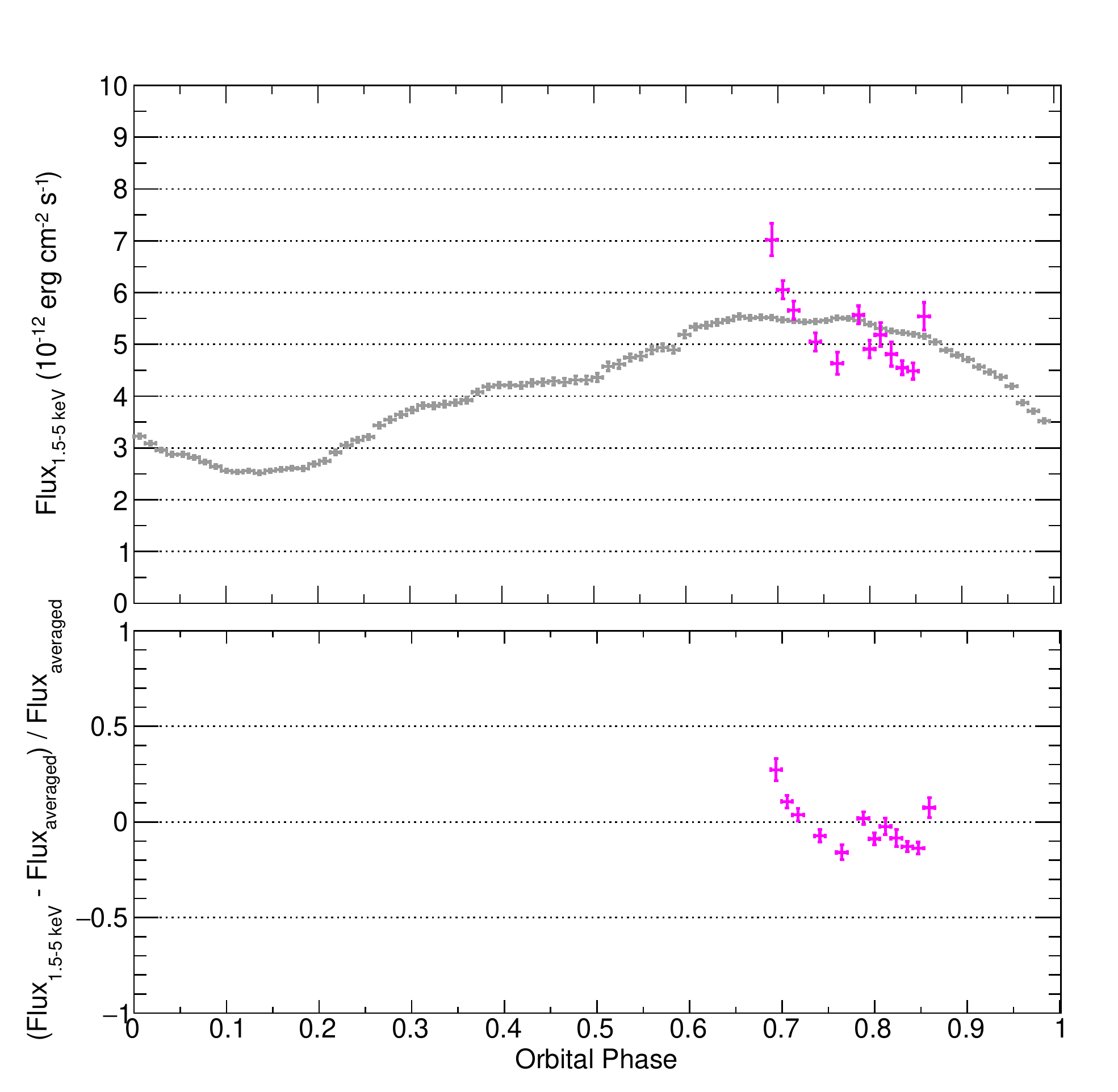}{0.4\textwidth}{(d) OBSID 4631010201--2}}
\caption{Orbital light curves for each \nic observation. The data points in grey are the same as those in black in the middle panel of Figure~\ref{fig_orbital_light_averaged}. 
The bottom of each figure shows the difference between the original and the running-averaged orbital light curves at each phase, normalized by the running-averaged one.}
\label{fig_each_orbital_light_curve_w_averaged_nustar}
\end{figure*}

\begin{figure}
\centering
\includegraphics[width = 0.45\textwidth]{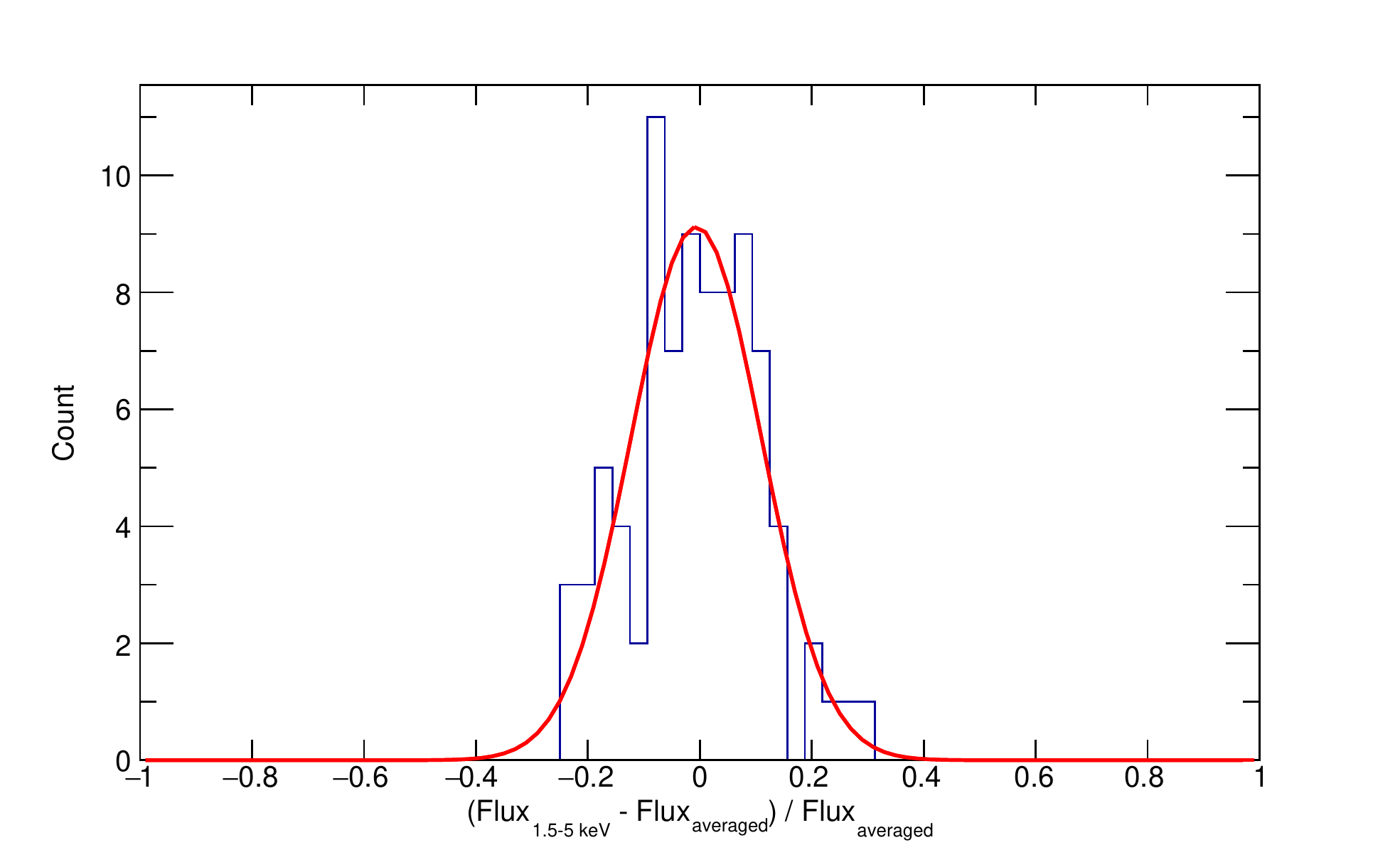}
\caption{The histogram of the difference between the original and the running-averaged flux for each time bin, normalized by the running-averaged one.
Here the energy range of the flux is 1.5--5 keV, and the four \nic observations are used. The window width of the running average is 70 ks.
The red line is the best fitting Gaussian function.}
\label{fig_hist_delfluxratio_NICER_all_averaged}
\end{figure}

To investigate the short-term variability inferred from Figure~\ref{fig_orbital_light_curve},
we show in Figure~\ref{fig_each_orbital_light_curve_w_averaged_nustar} the orbital light curve for each \nic observation with the running-average window width of 70~ks, i.e., the grey curves in Figure~\ref{fig_each_orbital_light_curve_w_averaged_nustar} are the same as the black one in the middle panel of Figure~\ref{fig_orbital_light_averaged}.
In all of the observations, it can be seen that the light curves cross the running-averaged one with typical time scales for the crossings of 2 to 8 bins, corresponding to about 10 to 30~ks.
Since the orbital modulation seems quite stable in the long term on time scales $\gtrsim 70$~ks, the differences between the actual and the running-averaged flux at each phase may be considered as flux fluctuations on top of the steady level. These variations are shown at the bottom of each panel in Figure~\ref{fig_each_orbital_light_curve_w_averaged_nustar}. We note that we normalized these differences by the running-averaged flux at each phase. Doing this, sudden flux changes can be seen more clearly. For instance, the figure panels show flux increases at $\phi = 0.2$--0.25 and $\phi = 0.45$--0.55 that last for 10--30~ks (b), wavy structures at $\phi = 0.3$--0.45 (b) and at $\phi = 0.6$--0.8 (c), and a fast drop from higher fluxes at $\phi\sim 0.7$ (d).
As explained in Appendix~\ref{sec_background_rate}, such structures are not artifacts induced by the background.
Finally, we present the histogram of the ratio of the flux difference for each time bin in Figure~\ref{fig_hist_delfluxratio_NICER_all_averaged},
which is well described by a Gaussian distribution with a standard deviation of $0.12 \pm 0.01$, also shown in the figure.
The histogram illustrates the typical flux differences with respect to the average behavior, with a maximum flux difference of $\sim 30$\%.

\begin{figure}
\centering
\includegraphics[width = 0.45\textwidth]{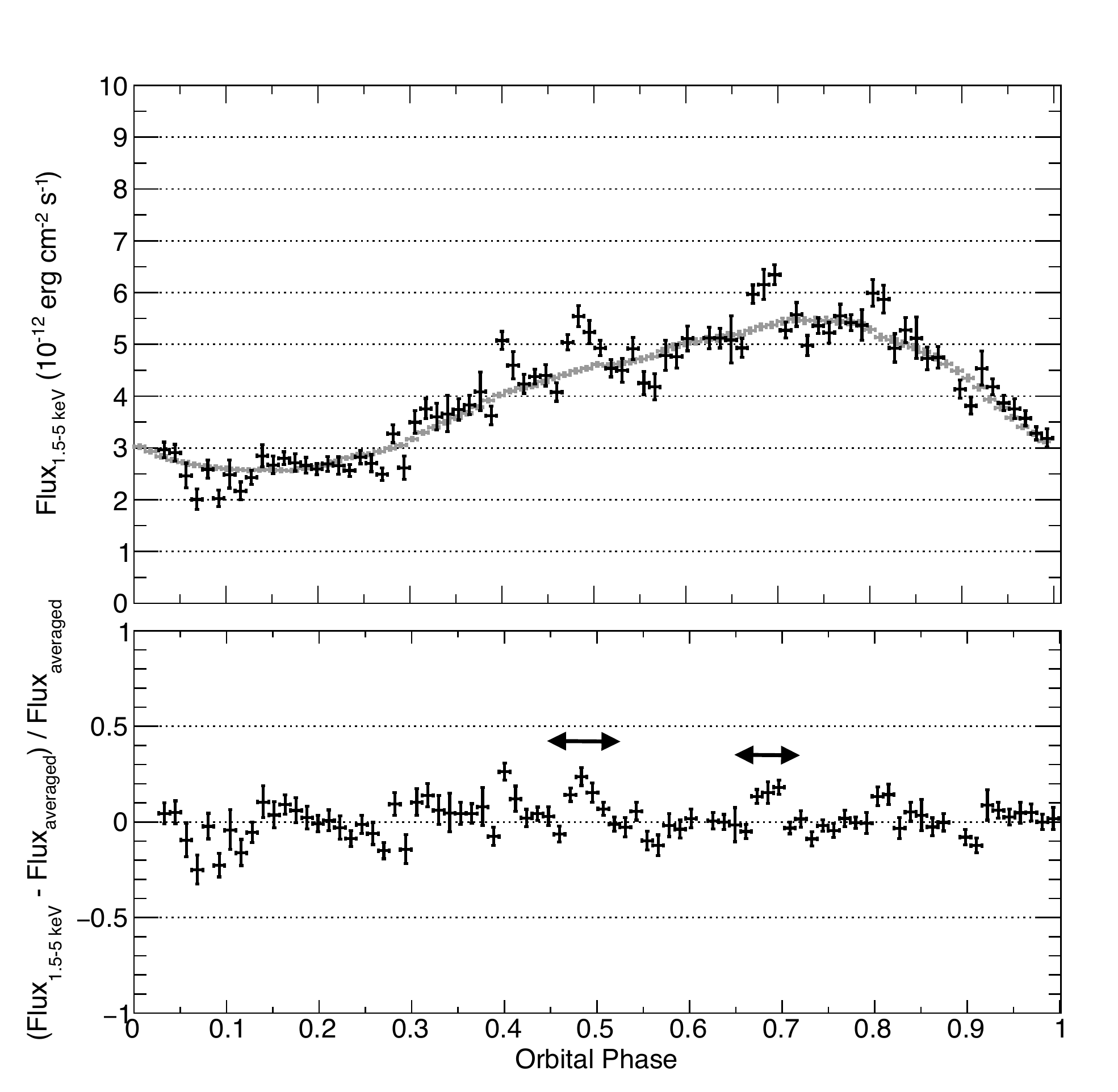}
\includegraphics[width = 0.45\textwidth]{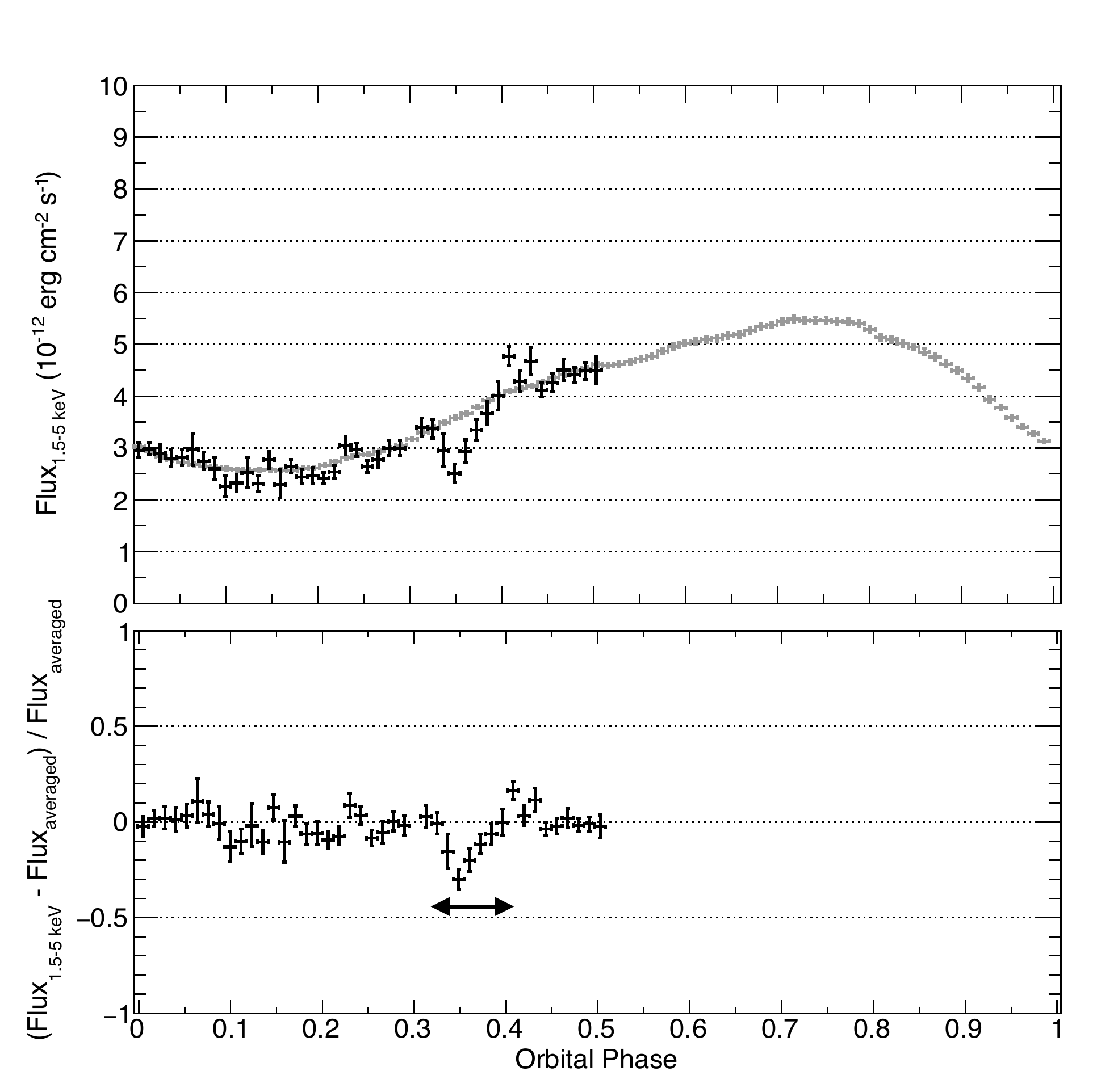}
\caption{The 1.5--5.0 keV orbital light curve of the \suzaku observation. Here the first orbit and the last half were analyzed separately. The grey data points are the same as those in the middle panel of Figure~\ref{fig_orbital_light_averaged}.}
\label{fig_nicer_orbital_light_curve_w_averaged_suzaku}
\end{figure}

We also examined the short-term variability present in the \suzaku observations.
Figure~\ref{fig_nicer_orbital_light_curve_w_averaged_suzaku} shows the first and the second orbits observed by \suzaku separately.
We compared the actual light curves with the running-averaged one.
In the top of Figure~\ref{fig_nicer_orbital_light_curve_w_averaged_suzaku}, sharp structures are seen at $\phi \simeq 0.5$ and $0.7$. We note that the latter was previously pointed out as an X-ray spike in \cite{kishishitalongterm2009}.
For the second orbit, a sudden flux change was found more clearly at $\phi\approx 0.35$, with a flux drop of $\approx 40$\% and a recovery back to the averaged flux within $\sim 20$~ks. All this is consistent with features found in the \nic observations.
We also extracted the events around $\phi \approx 0.35$ (i.e., the flux drop phase; specifically, in the time interval MET 243113028--243129028) and fitted the spectrum by setting the column density as a free parameter. We obtained a column density of $0.57 \pm 0.14 \times 10^{22}~\mathrm{cm}^{2}$, which is consistent with the value from the overall data. 
The photon index was $1.40\pm0.11$ and did not show a significant change either.
Provided that a column density of $\approx 4 \times 10^{22}~\mathrm{cm^2}$ is required to cause a flux drop of $\approx 40$\%, the short-term flux variability is be caused by an intrinsic change of the X-ray emission itself, and not by an optically thick medium temporarily covering the line of sight.

\subsection{Search for X-ray flares on time scales of $\sim 10$~s}\label{subsec_flare_search}

We also investigated variability on a much shorter time scale, of $\sim 10$~s, to search for X-ray flares similar to those detected in other gamma-ray binaries \citep[e.g.,][]{torresmagnetarlike2012} and to features seen in accreting systems \citep{Brocksopp1999,Grinberg2013}.
We produced a light curve with a bin width of 32~s and searched for time intervals with a flux at least five times larger than the average. Here the referred average flux is calculated for every good time interval, and the energy range is 1.5--5.0~keV. We note that the good time intervals less than 200 s were ignored in this analysis because the average fluxes have large statistical errors in these intervals.
As a result, we did not find any time intervals with such a large flux fluctuation.
We constrained the flare occurrence rate as less than 2.4 events per day with 95\% confidence level.
We note that we did not observe any flare-like events
even with a shorter time scale, e.g., a bin width of 100 ms or 1 s.

\subsection{Comparison with 2016 \nus data} \label{sec_nustar}

We also studied the orbital variability in the hard X-ray band using the 2016 \nus observation (OBSID: 30201034002). Because \nus is sensitive to hard X-rays above 3 keV and its energy coverage is different from \nic, the analysis presented in this subsection is complementary to the above results. 
We analyzed the \nus data in the range 3--10 keV, and compared the results with the \suzaku observation in 2007 in the same energy range.
The \nus data were already analyzed by \cite{Yoneda_2021} and \cite{Volkov2021}, and 
the data reduction and the spectral fitting used here are the same as in \cite{Yoneda_2021}.

\begin{figure}
\centering
\includegraphics[width = 0.45\textwidth]{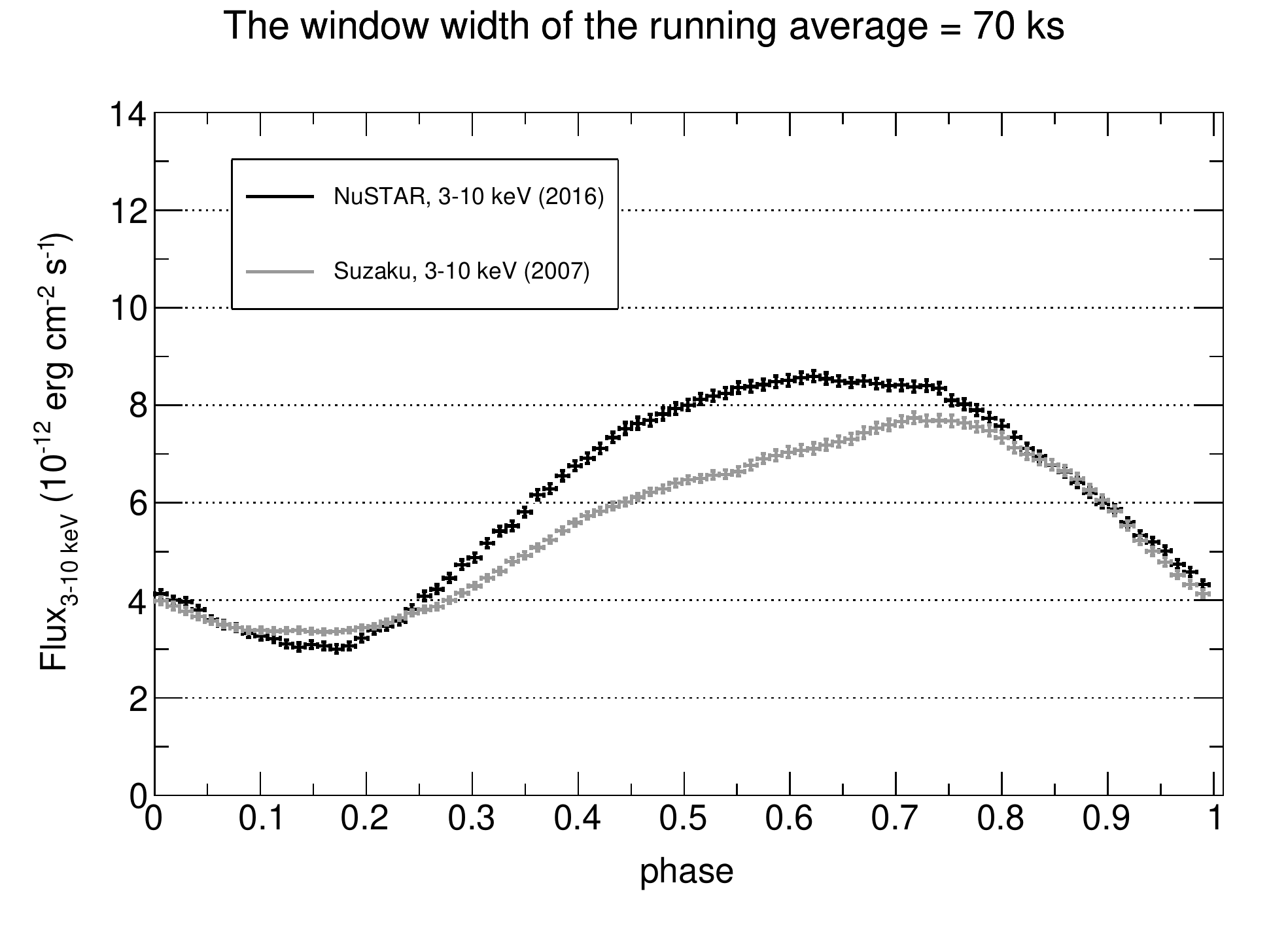}
\caption{Running-averaged orbital light curves of the \nus and \suzaku observations. The energy range in the spectral fitting is 3--10~keV. The window width of the running average is 70~ks.}
\label{fig_orbital_light_averaged_nustar}
\end{figure}

Figure~\ref{fig_orbital_light_averaged_nustar} shows the running-averaged orbital light curve obtained with the window width of \(70\rm\,ks\) from the \nus data.
Unlike the \nic observation (see Figure~\ref{fig_orbital_light_averaged}), the \nus data show an obvious flux difference from \suzaku, especially around $\phi$ of 0.3--0.7.
The maximum difference is 26\% at $\phi = 0.55$.
However, the orbital light curves still show a good consistency at $\phi$ of 0.8--1.1 with a maximum difference of less than 6\%.
This result points at two possibilities: the overall emission in 3--10 keV is somewhat less stable than at lower energies; and/or the activity of \ls was peculiar during the \nus observation.

\begin{figure}
\centering
\includegraphics[width = 0.45\textwidth]{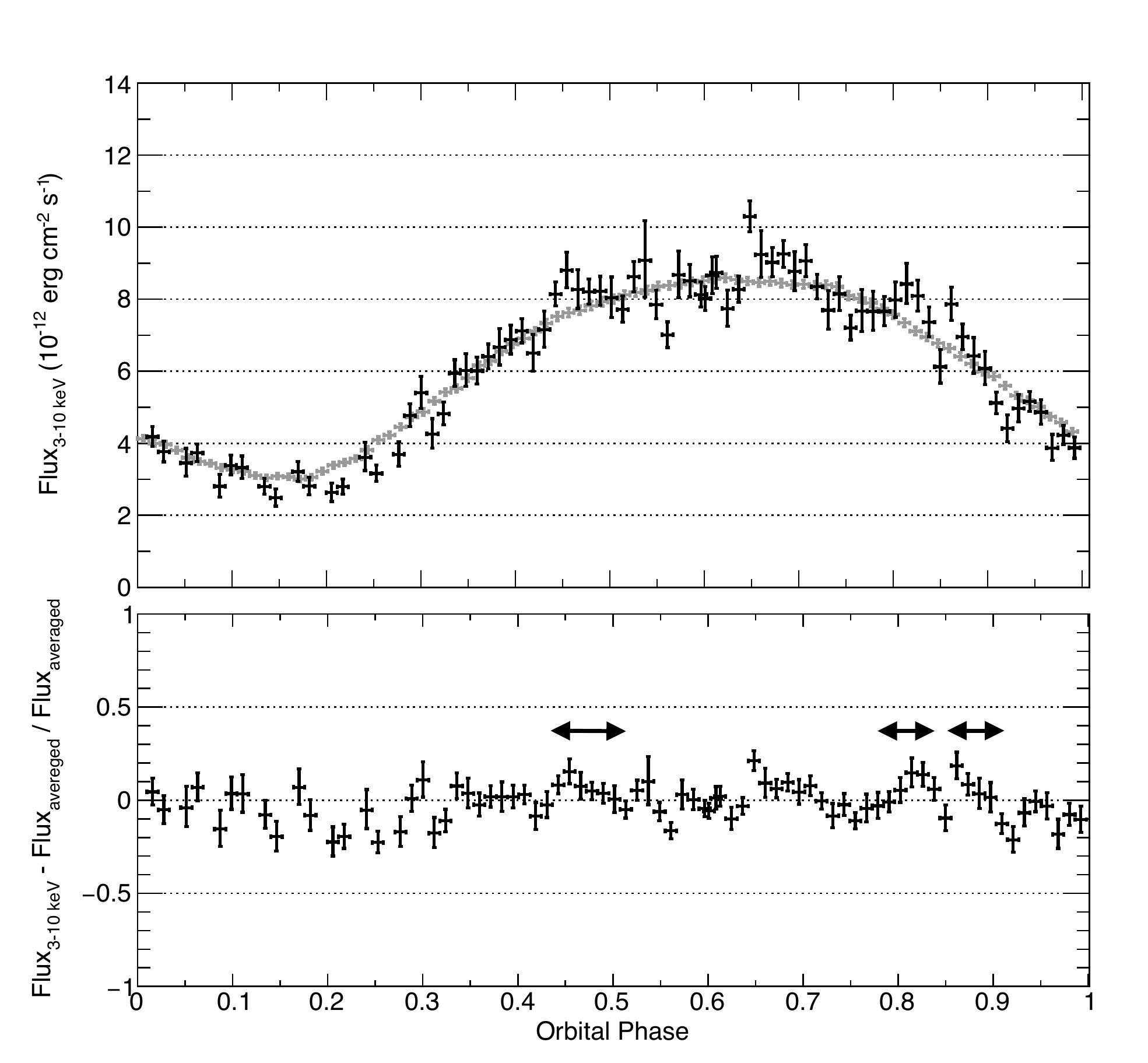}
\caption{Orbital light curve of the \nus observations in the range 3--10~keV. The grey data points are the same as the black ones in Figure~\ref{fig_orbital_light_averaged_nustar}.}
\label{fig_nustar_3_10keV}
\end{figure}

\begin{figure}
\centering
\includegraphics[width = 0.45\textwidth]{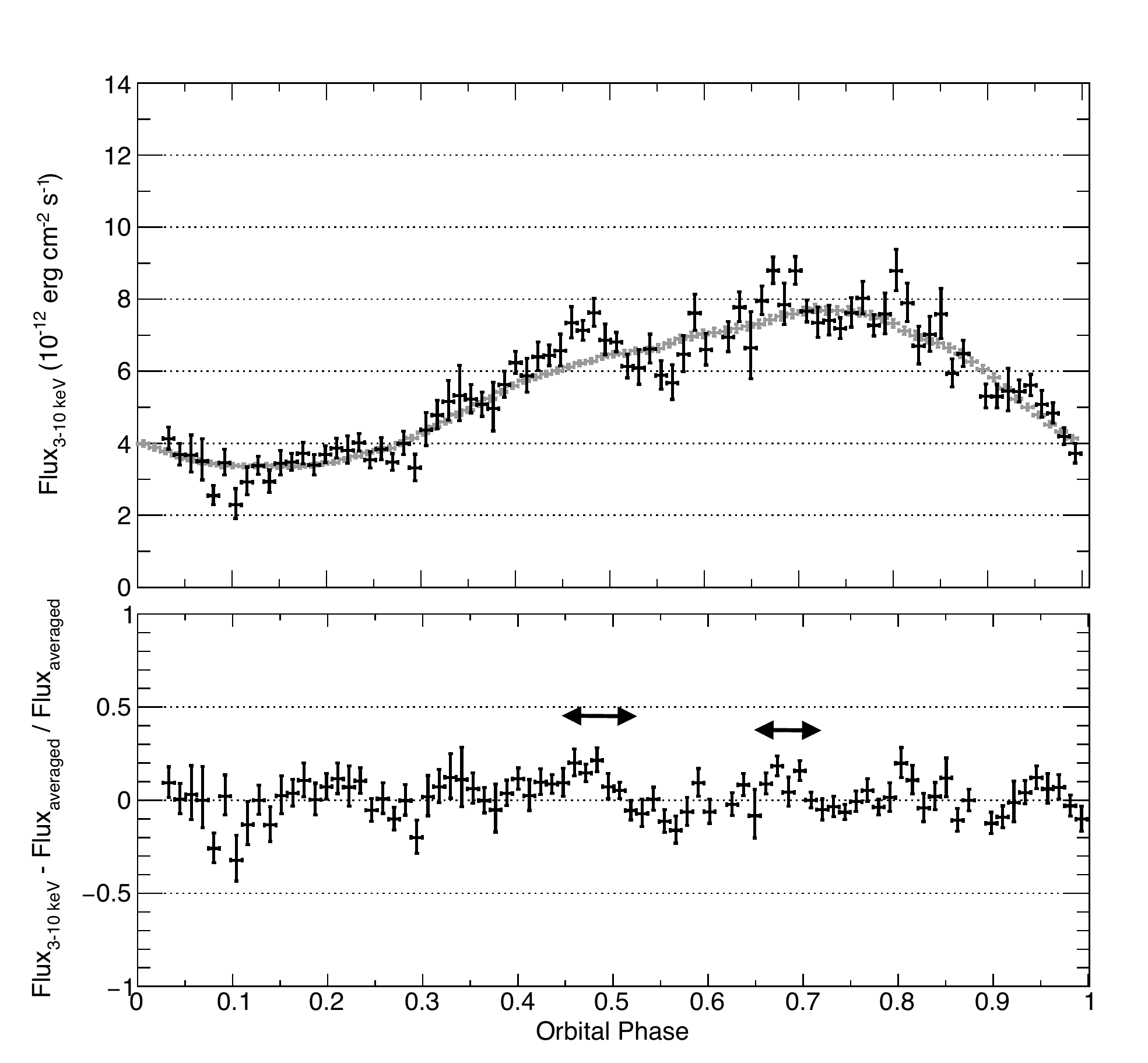}
\includegraphics[width = 0.45\textwidth]{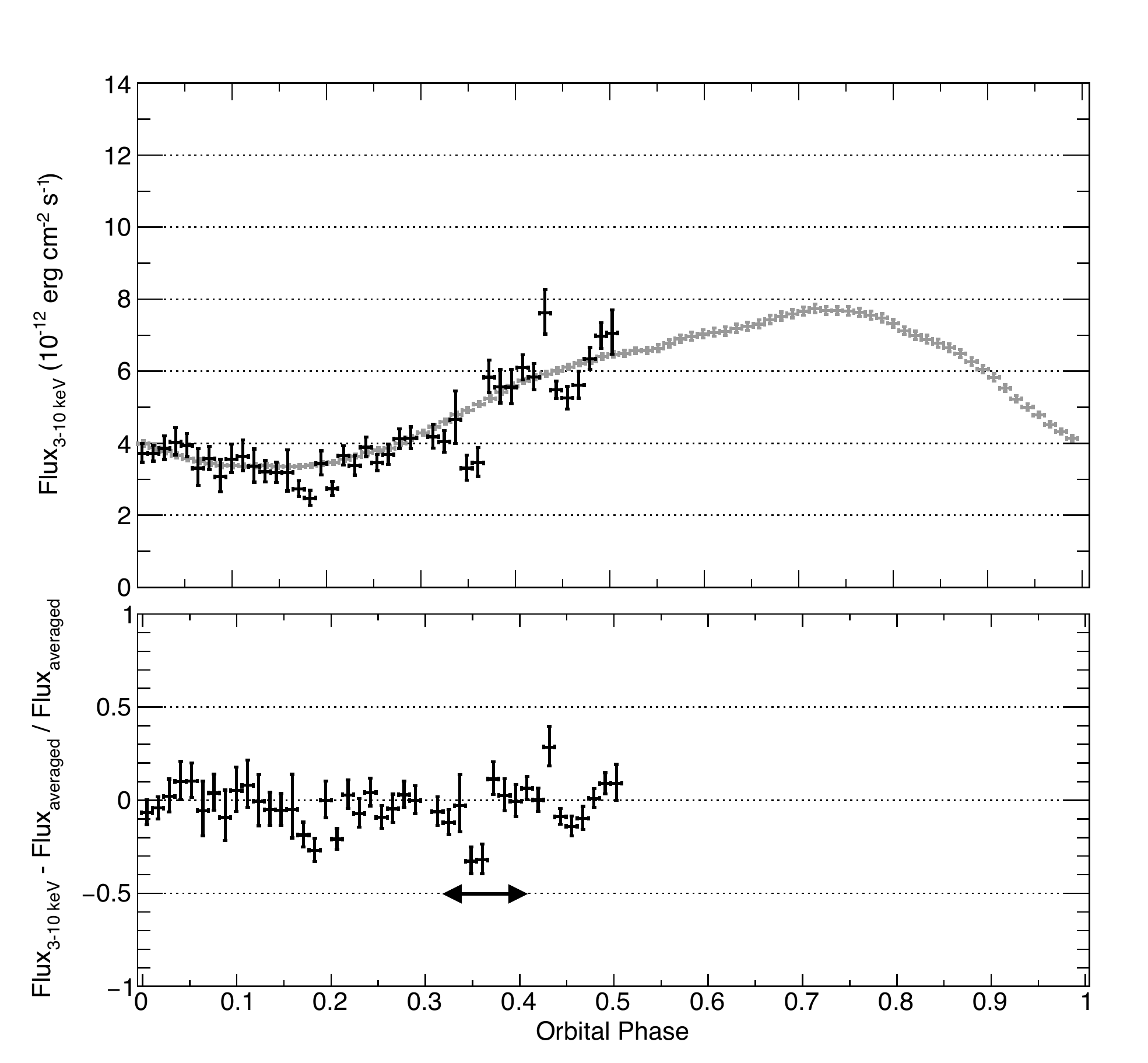}
\caption{Orbital light curve of the \suzaku observations in the range 3--10~keV. As before, the first and the second orbits were analyzed separately. The grey data points are the same as those in Figure~\ref{fig_orbital_light_averaged_nustar}.}
\label{fig_suzaku_3_10keV}
\end{figure}

\begin{figure}
\centering
\includegraphics[width = 0.45\textwidth]{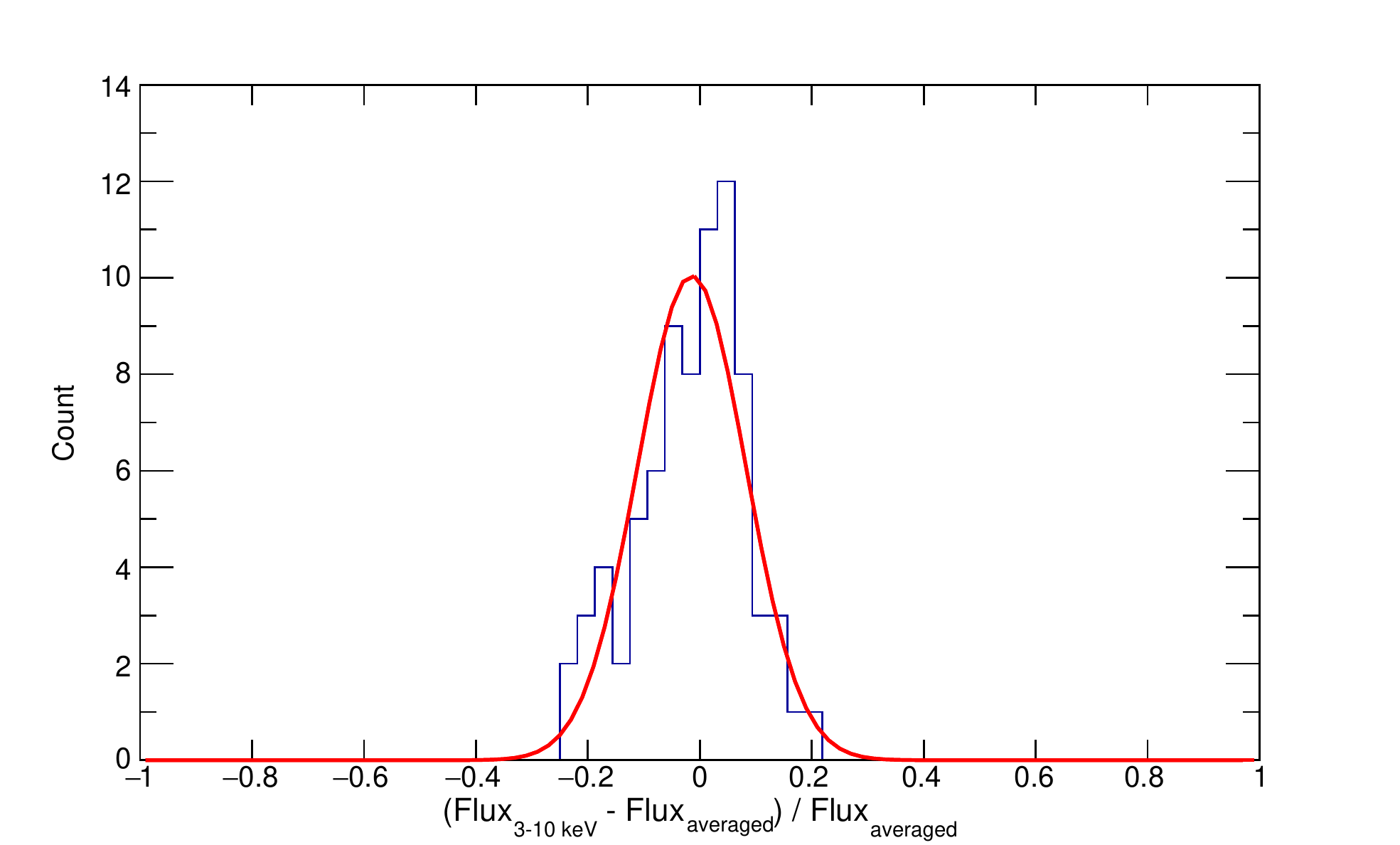}
\caption{The same as Figure~\ref{fig_hist_delfluxratio_NICER_all_averaged}, but for the 3--10 keV \nus data.}
\label{fig_hist_delfluxratio_3_10keV_nustar_nustar_averaged_Casares_ave_70ks}
\end{figure}

We also investigated the short-term variability in the range 3--10 keV.
Figure~\ref{fig_nustar_3_10keV} shows the orbital light curve of the \nus observation and its difference from its running-averaged curve. As in Figures~\ref{fig_each_orbital_light_curve_w_averaged_nustar} and \ref{fig_nicer_orbital_light_curve_w_averaged_suzaku}, Figure~\ref{fig_nustar_3_10keV} shows
evidence of variability with a time scale of a few tens of ks, as indicated with arrows in the figure.
We also show the 3--10~keV orbital light curve of the \suzaku observation in Figure~\ref{fig_suzaku_3_10keV}.
The arrows in the figure represent the same intervals shown in Figure~\ref{fig_nicer_orbital_light_curve_w_averaged_suzaku}.
As seen in the \nic, \suzaku, and \nus lightcurves, the overall short-term variability seems to be similar at these slightly higher energies at the precision level available.
As shown in Figure~\ref{fig_hist_delfluxratio_3_10keV_nustar_nustar_averaged_Casares_ave_70ks}, the standard deviation of the ratio of the flux difference with the 3--10 keV \nus data is $0.10 \pm 0.01$, consistent with the one obtained with the \nic observations.

\section{Discussion} \label{sec_discussion}

In the previous section, the orbital light curves from the 2018--2021 \nic observations and from the 2007 \suzaku observations have been compared. Although they are separated by $\sim 14$~years, the obtained fluxes in 1.5--5 keV are comparable over the whole orbit.
In particular, when using the running-averaged flux with a window width of 70~ks, the orbital curves obtained from the two instruments show remarkable consistency; for instance, the maximum flux difference is only $\approx 3$\% at $\phi \sim 0.1$--0.2 and $0.7$--0.8.
Furthermore, the \nic orbital light curve shows a short-term variability with a typical time scale of 10--30 ks.
Such a feature is also seen in the \suzaku light curve, and no simultaneous significant change in the column density is found in either of these observations.
These results suggest that the long-term stability of the orbital modulation, when averaging over time scales $\gtrsim 70$~ks, is a trace of an equilibrium state of the system, and the flux variability with $\sim$10 ks corresponds to perturbations of this equilibrium state. In what follows, we explore a promising astrophysical interpretation of this behavior based on the pulsar-stellar wind interaction scenario.

\subsection{Long-term stability of the orbital X-ray modulation} \label{subsec_discussion_reproducibility}

The nature of the compact objects in most of the gamma-ray binaries is a long-standing question, in particular for LS~5039, with popular scenarios being the microquasar jet and the pulsar wind models \citep[see, e.g.,][]{Paredes2000,Martocchia2005}.
As already discussed in \cite{kishishitalongterm2009}, the clock-like behavior of the soft X-ray emission of \ls would disfavor at least the standard microquasar scenario because accreting black holes usually show stochastic and chaotic time variability, and state transitions are often observed. Even if the X-ray emission is produced in the relativistic jet launched from a black hole, and not from the accretion flow, some variability originating in the accretion process may be expected. It cannot be discarded that the emitting region is far enough from the compact object for the memory of that variability to be lost, although the remarkable long-term stability of the running-averaged orbital light curve (Figures~\ref{fig_orbital_light_averaged} and \ref{fig_orbital_light_averaged_aragona}) and the lack of second-scale X-ray variability (\S\ref{subsec_flare_search}) tends to favor the stellar-pulsar wind scenario when compared to typical accreting sources.

In the standard pulsar wind scenario, particle acceleration takes place in the termination shock of the pulsar wind when interacting with the stellar wind. As far as these winds are approximately stable, the global structure of the interaction region should be relatively stable as well, as it is characterized by the balance between the two-wind ram pressures. The fact that orbital light curves separated by 14 years are very similar suggests that the properties of the pulsar and the stellar winds in \ls are indeed relatively stable over periods of at least 1--2 decades. It is worth noting that hydrodynamical simulations indicate that the stellar-pulsar wind interaction structure can quickly change its geometry along the orbit due to instability growth and other forms of flow perturbations \citep[see next section, and, e.g.,][]{BoschRamon2012,Paredes-Fortuny2015}, but the running-averaged light curves on scales $\gtrsim 70$~s are likely smoothing out those variations. Therefore, what these light curves manifest is probably the underlying long-term stability of the overall interaction structure.

\subsection{Short-term variability} \label{subsec_discussion_shortvaribility}

The short-term variability with a time scale of $\sim$10 ks can be also interpreted under the pulsar wind scenario.
As noted, if the colliding winds are relatively stable, the shock region formed due to ram pressure balance of these winds will be also relatively stable, at least when considering moderately long sub-orbital timescales of $\sim 1$~day (recall that the period of \ls is $\approx 3.9$~days).
In this context, the short-term flux variability should be caused by some effects yielding deviations from the equilibrium state of the two-wind interaction structure, as for instance inhomogeneities or time variability in the wind properties, instabilities in the shock region, and so on. For example, the stellar winds from massive stars are known to have irregularities or clumps in their density and velocity distributions \citep{Runacres2002}.
In addition, the orbital curve of the O-star radial velocity shows relatively large variations \citep{casarespossible2005}, which may suggest the presence of complex stellar wind structures in \ls. 
We note that a presence of a large clump is also discussed in the gamma-ray binary \psrb \citep{Hare2019}.

\begin{figure*}
\centering
\includegraphics[width = 0.7\textwidth]{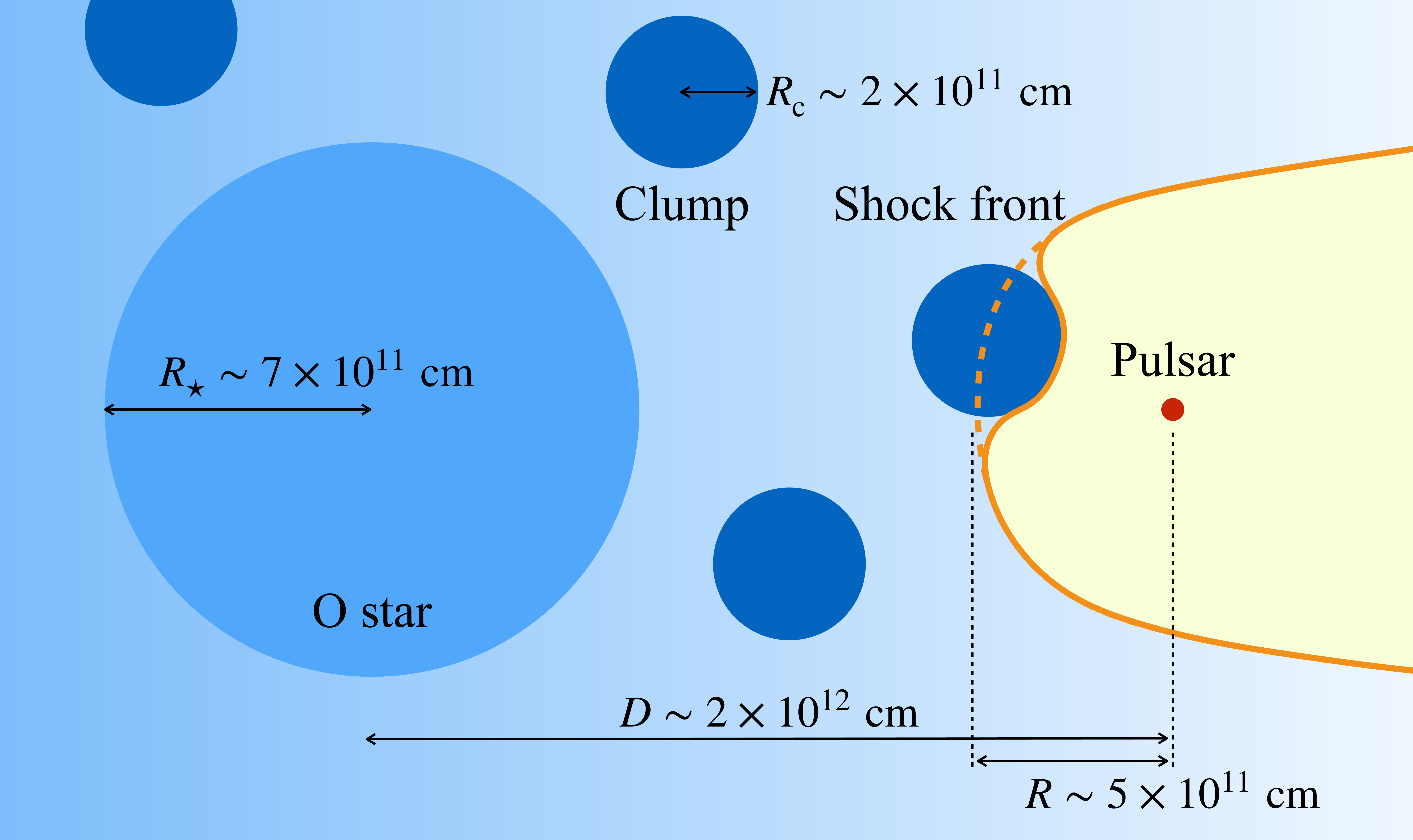}
\caption{A schematic of the interaction between the stellar clump and the pulsar winds.}
\label{fig_clumpy_pulsar_wind}
\end{figure*}

When inhomogeneities of different scales are present in the stellar wind, it is expected that the interaction structure will vary in space and time \citep{BoschRamon2013,Paredes-Fortuny2015,Kefala2022}, resulting in the observed short-term variability. Following this idea, we discuss in what follows the characteristic values of time/size scale and frequency \citep[e.g.,][]{BoschRamon2013}. By comparing them with the observed values, we examine whether the pulsar wind scenario with a clumpy stellar wind is supported by the variability patterns found in X-rays. We show a schematic of the discussed scenario in Figure~\ref{fig_clumpy_pulsar_wind}.

We note that \cite{BoschRamon2011,BoschRamon2012, takata2014} discussed the formation of a strong termination shock in the opposite direction of the companion star, with particle acceleration potentially taking place even further downstream  \citep{Zabalza2013}. However, hereafter we focus on the shock region between the two stars, which is expected to be more stable than the emitting regions outside the binary system. We note that a non-negligible amount of X-rays and radiation at other wavelengths may come from those further locations, but it is enough for our purposes if only a significant X-ray component originates in the intrabinary region.

\subsubsection{The clump-pulsar wind interaction scenario} \label{subsubsec_clump_pulsar_interaction}

 When a clump from the stellar wind interacts with the pulsar wind,
the two-wind interaction structure will get deformed until the pulsar wind destroys the clump. This time scale ($t_{\mathrm{c}}$) can be estimated as \citep{BoschRamon2013}
\begin{align}
t_{\mathrm{c}} \approx \frac{f^{-1/2} R_{\mathrm{c}}}{V_{\mathrm{w}}}~,
\end{align}
where $R_{\mathrm{c}}$ and $V_{\mathrm{w}}$ are the radius of the clump and the stellar wind velocity, respectively, and $f$ is the stellar wind-to-clump density ratio, for which we assume $\sim 0.1$--0.01. 
Since the acceleration and cooling time scales (synchrotron, inverse Compton and escape/adiabatic losses) are much shorter than $t_c$ (see \citealt{khangulyan2008}, \citealt{BoschRamon2013}, \citealt{Yoneda_2021}), 
the duration of the X-ray variation induced by the clump presence is mainly determined by $t_c$.
By equating $t_{\rm c}$ to the observed short-term variability time scale, the clump size can be constrained to
\begin{align}
\begin{split}
R_{\mathrm{c}} &\approx 2\times10^{11}~\mathrm{cm} \\
&\times \left(\frac{f}{0.01}\right)^{1/2}
\left(\frac{t_{\mathrm{c}}}{10~\mathrm{ks}}\right)
\left(\frac{V_{\mathrm{w}}}{2000~\mathrm{km/s}}\right)~.
\label{rcl}
\end{split}
\end{align}
We note that the clump does not necessarily have its origin in the instabilities associated to wind formation; relatively large inhomogeneities may also be present in the stellar wind due to other factors intrinsic to the star (e.g., rotation, pulsations, etc.) or even related to the pulsar presence (e.g., through its gravity or its radiation).

One can compare $R_{\mathrm{c}}$ with two relevant quantities, the size of the stellar-pulsar wind shock region ($R$), and the radius of the companion star $R_\star$.
Regarding $R$, assuming that it is comparable to the distance between the shock and the pulsar it can be derived as
\begin{align}
R &= \frac{\eta^{1/2}}{1 + \eta^{1/2}} D \\
&\simeq 5\times10^{11}~\mathrm{cm} \times
\left(\frac{\eta}{0.09}\right)^{1/2}
\left(\frac{D}{70~\mathrm{lt\text{-}sec}}\right)~,
\end{align}
where $D$ is the binary system separation, and $\eta$ the momentum flux ratio between the pulsar and stellar winds:
\begin{align}
\eta = \frac{L_{\mathrm{sd}}}{\dot{M}_{\mathrm{w}} V_{\mathrm{w}} c}~,
\end{align}
where $\dot{M}_{\mathrm{w}}$ is the mass-loss rate of the stellar wind and $L_{\mathrm{sd}}$ the spin-down luminosity of the pulsar (see Table~\ref{tab_parameters}), and $c$ the speed of light.
Since $R_c \sim 0.4 R$, in this scenario the X-ray variability would be caused by a relatively large clump comparable in size to the X-ray emitting region considered.
Regarding $R_\star$, \ls harbors an O star, and $R_\star$ has been estimated to be $\approx 9.3\,R_\odot \approx 7\times 10^{11}$~cm \citep{casarespossible2005}. This suggests that the clump impacting the two-wind interaction structure should be rather large, with a radius $\sim 0.3\,R_\star$.

Now, theoretically, it is uncertain whether stellar wind clumps can have such a large size, although as noted some stellar wind inhomogeneities may in principle be large enough (see \S3.3 in \citealt{BoschRamon2013} for details). Nevertheless, it must taken into account that the chaotic arrival of smaller clumps can lead to non-trivial behaviors of the X-ray light curve, with short-term variability features and their rates being dependent on the stellar wind clumping properties. 
In particular, \cite{Kefala2022} results indicate that a few smaller clumps (compared to $R_{\rm c}$ above) can impact simultaneously on the two-wind interaction region, inducing an evolution time of the perturbation longer than in the case of only one clump.
We note that stellar wind inhomogeneities large enough to engulf the two-wind interaction region, even under moderate density contrasts (say a factor of $\sim 2$), may also induce variations in the structure of the two-wind interaction region. These variations would have a characteristic time scale of a few $\times R/V_{\mathrm{w}} \sim$ 10 ks and could still significantly impact the light curve properties. 

The impact of clumps on the X-ray short-term variability might be especially pronounced if a significant fraction of the detected X-ray emission is produced by shocked pulsar wind moving relativistically towards the observer \citep[see, e.g.,][]{Bogovalov2008}. In this case, because of induced changes of the Doppler boosting factor, even a relatively small perturbation of the plasma bulk speed may cause a considerable increase or decrease of the emission towards the observer. It is worth noting that this effect may provide a simple explanation for the fact that close to the superior conjunction phases the X-ray light curve seems to be less prone to the short-term variability, as during those periods the relativistically moving plasma is expected to propagate away from the observer. 

\begin{table}[!htbp]
  \begin{center}
  \caption{The typical parameter values for \ls, referring to \cite{casarespossible2005}, \cite{collmarls2014}.}
\label{tab_parameters}
    \begin{tabular}{c c} 
parameter & value \\ \hline
$\dot{M}_{\mathrm{w}}$ & $3 \times 10^{-7}~M_\odot~\mathrm{yr}^{-1}$\\
$v_{\mathrm{w}}$ & $2000~\mathrm{km~s^{-1}}$\\
$D$ & 70 lt-sec\\
${L_{\mathrm{sd}}}^a$ & $10^{37}~\mathrm{erg~s^{-1}}$\\
\hline
    \end{tabular}
\\$^a$ For a non-thermal luminosity $\sim 10$\% of $L_{\rm sd}$.
\end{center}
\end{table}

Using the fact that the column density does not change significantly, with an uncertainty $\lesssim 10^{22}$~cm$^{-2}$, we can constrain the clump over-density factor $f$ as for certain types of clumps their presence should affect the column density.
Using the source parameters, the averaged wind number density $n_{\mathrm{w}}$ can be calculated as
\begin{align}
n_{\mathrm{w}} = \frac{\dot{M}_\mathrm{w}}{4 \pi D^2 V_{\mathrm{w}} m_p} \simeq 1\times10^{9}~\mathrm{cm^{-3}},
\end{align}
and the column density of the clump can be estimated as
\begin{align}
n_{\mathrm{w}} f^{-1/2} t_{\mathrm{c}} V_{\mathrm{w}}\lesssim 10^{22}~\mathrm{cm^{-2}}~.
\end{align}
It yields in 
\begin{align}
\begin{split}
f &\gtrsim 10^{-2} \\
&\times \left(\frac{t_{\mathrm{c}}}{10~\mathrm{ks}}\right)^{2}
\left(\frac{D}{70~\mathrm{lt\text{-}sec}}\right)^{-4}
\left(\frac{\dot{M}_\mathrm{w}}{3\times10^{-7}~M_\odot~\mathrm{yr}^{-1}}\right)^{2}~.
\end{split}
\end{align}
This result does provide an informative constraint on $f$, which disfavors values of $f$ in the range $\lesssim 0.01$.

Finally, we crudely estimate the frequency of the clump-pulsar wind interactions assuming the clump size obtained from Eq.~(\ref{rcl}).
The mass of the clump $M_{\mathrm{c}}$ can be estimated as
\begin{align}
M_{\mathrm{c}} = \frac{4\pi}{3} R_{\mathrm{c}}^3 f^{-1} n_{\mathrm{w}} m_{\mathrm{p}} \simeq 6\times10^{21}\times\left(\frac{f}{0.01}\right)^{1/2}~\mathrm{g}~.
\end{align}
We can also calculate the interaction frequency $F_{\mathrm{c}}$ by assuming that all the wind mass is in these clumps:
\begin{align}
F_{\mathrm{c}} &= \frac{R^2}{4 D^2} \frac{\dot{M}_\mathrm{w}}{M_{\mathrm{c}}} \\
&\sim 4 \times 10^{-5}~\mathrm{Hz}~.
\end{align}
This is actually an upper-limit to the interaction frequency, as it is very unlikely that all the wind mass is in the form of such large clumps. Nevertheless, if a significant fraction of the stellar wind is in the form of relatively large clumps, the interaction rate may be $\sim 10$ clump-pulsar wind interactions per orbit, which roughly matches the observed short-term behavior of the orbital light curves. It is worth noting that the more quantitative analysis derived by \cite{Kefala2022}, which adopted a more realistic clump mass distribution for the stellar wind and considered simultaneously many clumps, also predicts similar rates to those obtained here.

\subsubsection{Recurrent X-ray brightening at specific orbital phases?} \label{subsec_other_remarks}

\cite{kishishitalongterm2009} reported an X-ray flux increase around the inferior conjunction in both \chandra and \suzaku observations.
Although the two data are separated by 3 years, they lie at the same orbital phase, which raises a possibility that some of X-ray brightenings may occur at the same orbital phases and are repetitive.
However, \cite{Yoneda_2021} analyzed the \nus data and reported that an X-ray increase at the inferior conjunction was not found. 
\cite{Yoneda_2021} discussed several interpretations, namely that the sharp feature could be generated at the same orbital phase in every orbit but only in the soft X-ray band ($<$ 3 keV), or that the result from \chandra is just the increase of the overall X-ray emission since it covers only a small fraction of the orbit.
Here, with the addition of the \nic observations, we further discuss the repetition of the X-ray brightening.

In the orbital light curves from \nic, \suzaku, and \nus (see Figures~\ref{fig_each_orbital_light_curve_w_averaged_nustar}, \ref{fig_nicer_orbital_light_curve_w_averaged_suzaku}, and \ref{fig_nustar_3_10keV}), some sharp features appear at similar phases in some of the observations.
For example, the \suzaku 1st orbit, \nic (OBSID:2030170102--4), and \nus observations show similar flux increases at $\phi\sim$0.5.
Also, at $\phi\sim 0.8$, the \suzaku 1st orbit, \nic (OBSID: 4631010101--2) and \nus observations show a sign of similar flux increases.
However, in the \suzaku 2nd orbit and the \nic 4th observation (OBSID4631010201--2), the X-ray brightening was not clearly seen at $\phi \sim$ 0.5 and 0.8, respectively, and 
thus the X-ray brightening does not seem to repeat itself always at the same orbital phase, but one can still claim that there are some orbital phases in which flux variations often occur. While the reason for this is not clear, the stellar wind could have a global pattern that tends to perturb the X-ray emission at similar orbital phases, for instance due to the stellar wind being affected by the gravity or the radiation from the pulsar. In such a case, the stellar wind inhomogeneities or anisotropies would not be completely chaotic, and their behavior would be leaving an imprint in the X-rays \citep[in][a link was already suggested between radial velocity curves and short-term variability found in {\it RXTE} data]{BoschRamon2005}.

Although an analysis of the multiwavelength behavior of LS~5039 is beyond the scope of this work, it is worth noting that flux spikes were also found above 100~GeV at similar orbital phases \citep{Aharonian2006b}. If these flux spikes have the same origin as the X-ray flux increases reported in this work, combining these two bands should allow the  study of the emission region in more detail (e.g., the X-ray/TeV flux ratio can inform on the magnetic energy density in the synchrotron plus inverse Compton scenario). For instance, \cite{BoschRamon2013}, \cite{delacita2017} or \cite{Kefala2022} calculated in detail the SED accounting for a clumpy stellar wind and showed that its spectral shape varies depending on the magnetization parameter in the shocked pulsar wind. Therefore, a future coordinated multiwavelength observation in X-ray and TeV bands with a large effective area, for instance, FORCE \citep{FORCE2016}, HEX-P \citep{HEXP2019}, and CTA \citep{CTA2011}, is important to reveal the origin of the short-term variability and the physical environment around the mysterious compact object in this system.

Regarding the general multiwavelength behavior of this source, we note that in the MeV--GeV and the 10--100 GeV regimes, the lightcurves obtained folding in phase long exposure observations by {\it COMPTEL} and {\it Fermi} (so averaging different orbital information), may suggest a smooth transition between the synchrotron and the inverse Compton components \citep{collmarls2014,Chang2016,Yoneda_2021}. Unfortunately, the poor statistics at those energies do not allow for studying sub-orbital timescale variabilities.

\subsection{A comment on the 9-second pulsation candidate} \label{subsec_pulse_search}

Signs of a 9-second pulsation in hard X-rays were reported by \cite{yoneda2020}, which might be another evidence for the presence of a neutron star in \ls. The inferred period and its derivative would suggest that a compact star is a magnetar rather than a normal pulsar, which potentially has an impact on the interpretation of this system. However, the significance of the pulse candidate is only at $\sim 3 \sigma$ for the \suzaku observations and much lower for those of \nus \citep{Volkov2021}, and confirmation is needed for the pulse detection.
Thus, we performed a pulsation search in \nic data focusing on the 9-second pulse candidate.

We followed the timing analysis performed in \cite{yoneda2020}.
We divided the data into subsets, calculated the $\mathrm{Z^2}$ statistics \citep{deJager1989}, and merged them.
The time duration of each subset varies from 3000 s to 12000 s with a step of 1000~s.
The timing analysis was performed for each of the four \nic observations.

While a pulsation with at least a 10\% pulse fraction can be detected with this method considering the count rates and background, no significant peak was detected in the range from 7 to 11 seconds.
We note however that the energy range used here is 1.0--5.0 keV, which is very different from the one in the previous report (10--30 keV).
More detailed analysis, that is, pulse search with demodulation of the binary motion, is out of the scope of this paper and is left for future work.

\section{Conclusion} \label{sec_conclusion}
We studied the short-term variability of the soft X-ray emission of \ls, one of the brightest gamma-ray binary systems in the galaxy. 
By analyzing the \nic observations from 2018 to 2021 and the \suzaku observations in 2007, we found that (1) the running-averaged orbital light curves from the two instruments show remarkable consistency with a window width of $\gtrsim 70$~ks,
(2) short-term variability on top of the orbital modulation was found by \nic and \suzaku, its time scale being $\sim 10$~ks. The standard deviation of the difference from the average flux obtained from \nic data is 12\%, and the maximum flux change was $\sim$30\%. Similar results described in (1) and (2) were also obtained when comparing with \nus data at 3--10~keV.
(3) No significant change in the column density was observed when the X-ray flux varied.
We interpret that the observed short-term variability is caused by the effect of relatively large clumps from the wind of the companion O star impacting on the two-wind interaction region. The characteristic values for the event duration, frequency, and clump size derived in that scenario are approximately consistent with observations, supporting the existence of an intrabinary shock formed by a pulsar, and a rather inhomogeneous stellar wind. 

\begin{acknowledgments}
The authors would like to thank the anonymous referee who provided a constructive report.
The authors would also like to thank Kazuo Makishima for useful comments.
We acknowledge support from JSPS KAKENHI grant numbers 20H00153, and 20K22355, and by RIKEN Incentive Research Projects.
V.B-R. acknowledges financial support from the State Agency for Research of the Spanish Ministry of Science and Innovation under grant PID2019-105510GB-C31AEI/10.13039/501100011033/ and through the ''Unit of Excellence Mar\'ia de Maeztu 2020-2023'' award to the Institute of Cosmos Sciences (CEX2019-000918-M). V.B-R. is Correspondent Researcher of CONICET, Argentina, at the IAR.
ZW acknowledges support by NASA under award number 80GSFC21M0002.
\end{acknowledgments}

\appendix
\section{Uncertainty of Orbital Parameter} \label{sec_unc_orbparam}

\begin{figure}
\centering
\includegraphics[width = 0.45\textwidth]{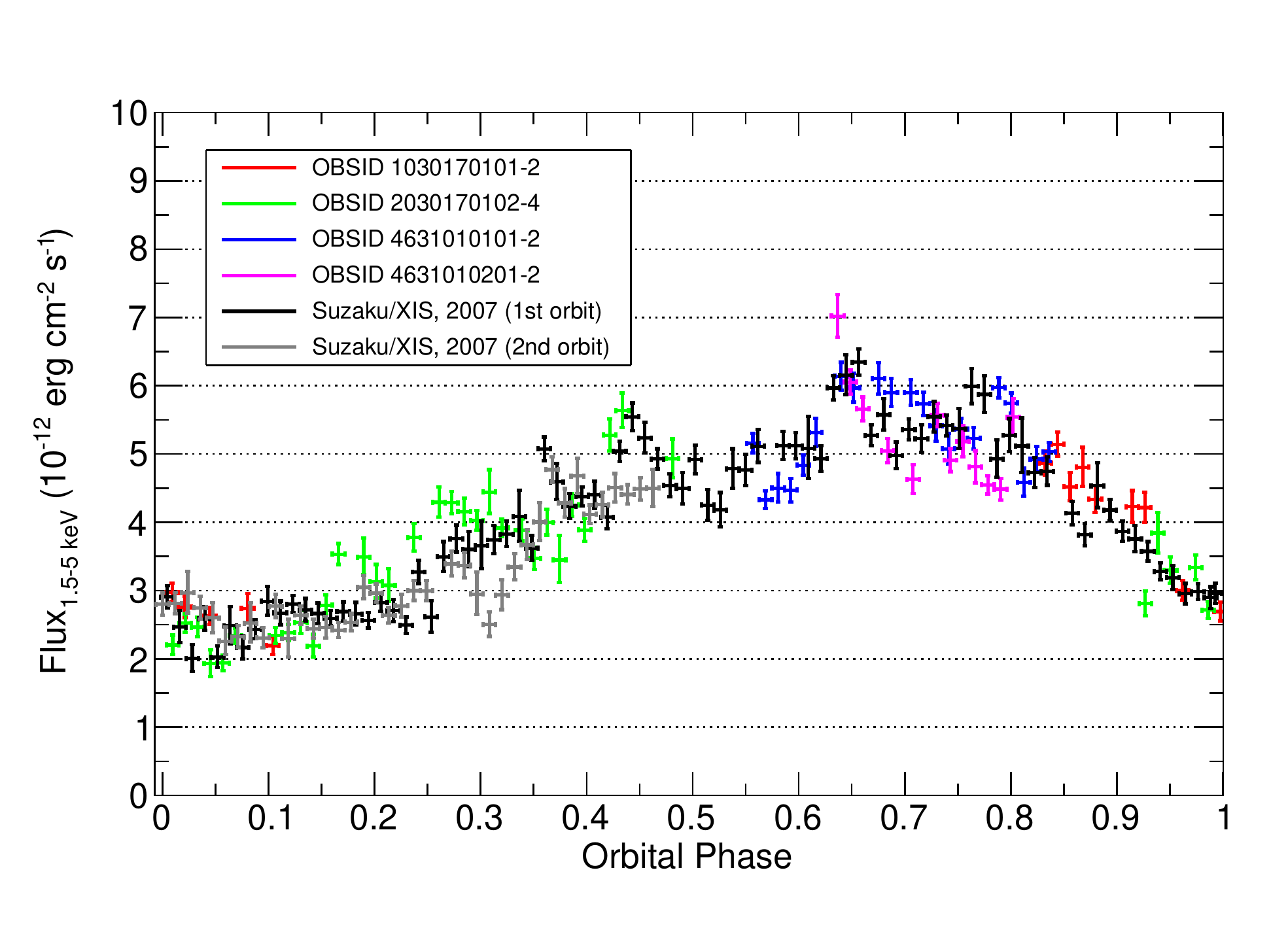}
\includegraphics[width = 0.45\textwidth]{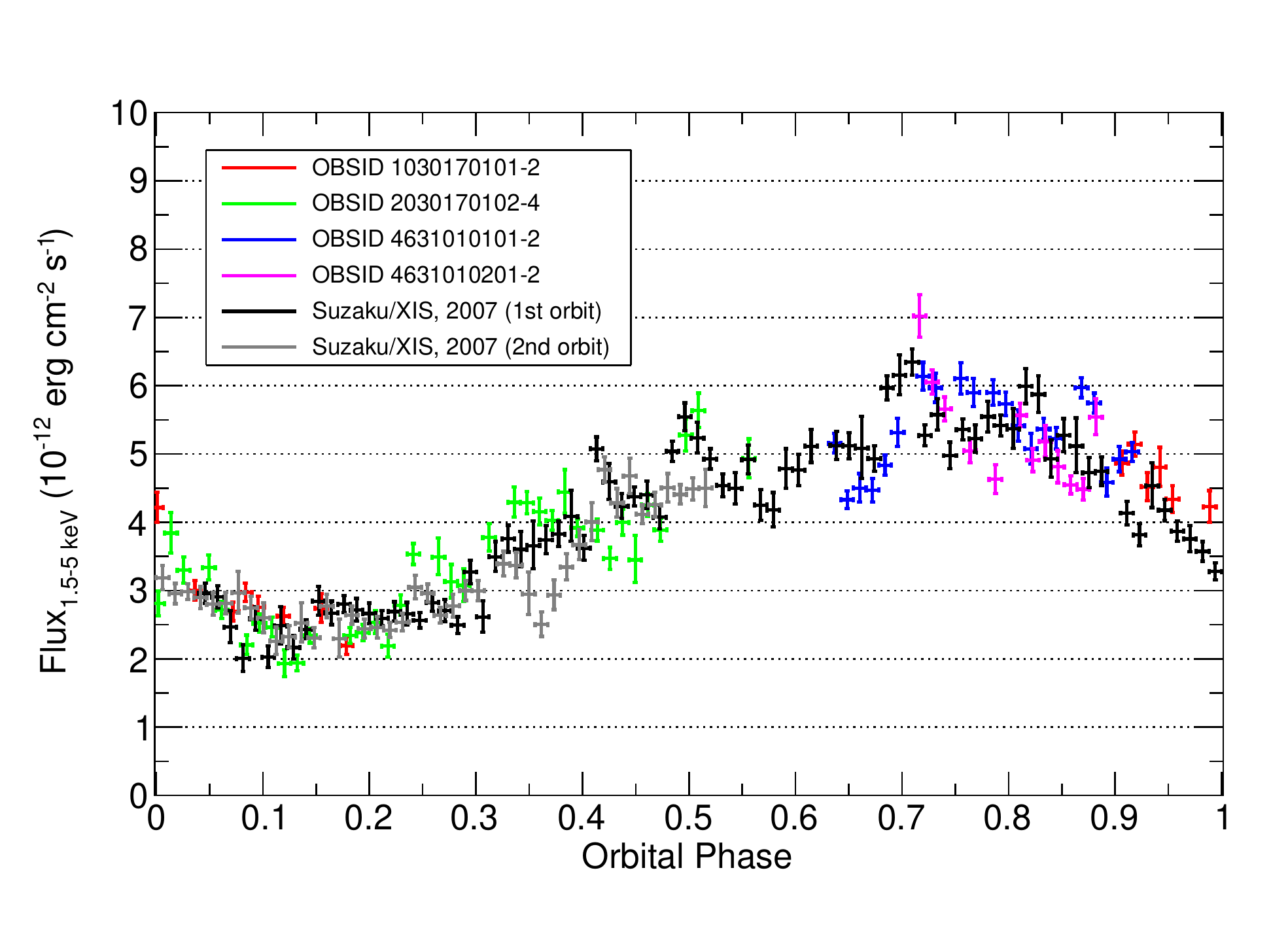}
\caption{The same as Figure~\ref{fig_orbital_light_curve}, but different orbital parameters were adopted. The left and right corresponed to \cite{aragonaorbits2009} and \cite{Sarty2011}, respectively.}
\label{fig_orbital_light_curve_diff_param}
\end{figure}

The orbital parameter sets have also been obtained by \cite{aragonaorbits2009, Sarty2011}.
Figure~\ref{fig_orbital_light_curve_diff_param} shows the orbital light curve adopting these orbital solutions.
The absolute uncertainty of the orbital phase due to different orbital solutions is 0.08 at the maximum.
The relative uncertainty between the \suzaku and the \nic observations is 0.027 at the maximum.
This latter uncertainty is important when we discuss the recurrent pattern in the orbital light curve. It is small and does not affect the argument in Section~\ref{subsec_other_remarks};
a short spike at $\phi\sim0.5$ can be also seen in Figure~\ref{fig_orbital_light_curve_diff_param}.
For the long-term stability, 
we can see the effect of different orbital solutions on the running-averaged curve in Figures~\ref{fig_orbital_light_averaged} and \ref{fig_orbital_light_averaged_aragona}.
With a window width of 70 ks, 
the \nic and \suzaku light curves are closer at $\phi = 0.7 - 1.1$ when using \cite{aragonaorbits2009}, i.e., the maximum difference between them is just 8\% with \cite{aragonaorbits2009} while it is 16\% with \cite{casarespossible2005}.
Thus, a few percent of the flux difference in the orbital light curves could be ascribed to the uncertainty in the orbital solutions. However, the maximum flux difference over the entire orbit is less than $\sim$20\% with any orbital solution, and we conclude that the argument about the long-term stability is robust regardless of the orbital solution uncertainty.

\section{Background Count Rate} \label{sec_background_rate}
\begin{figure*}
\gridline{\fig{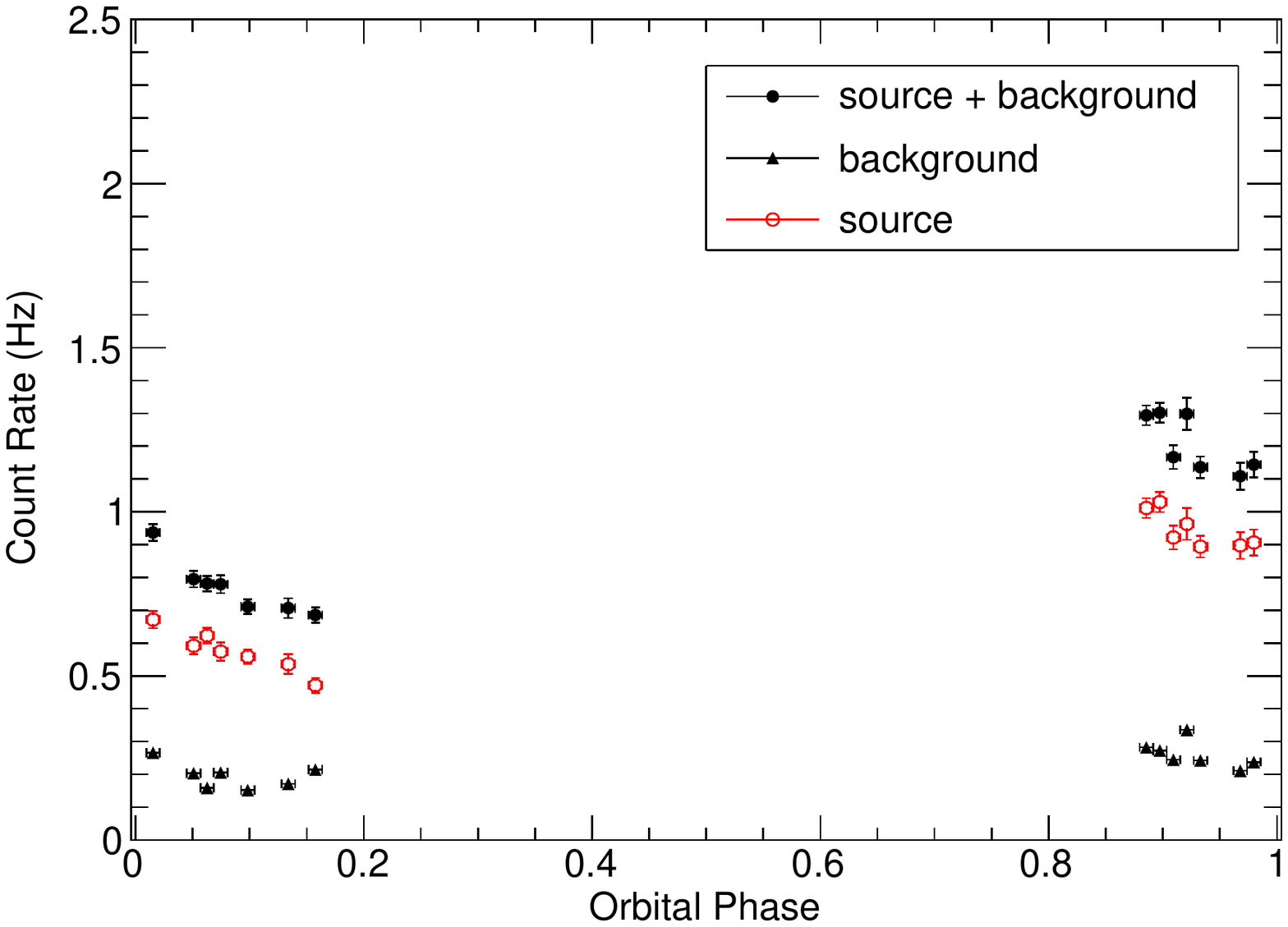}{0.4\textwidth}{(a) OBSID 1030170101--2}
          \fig{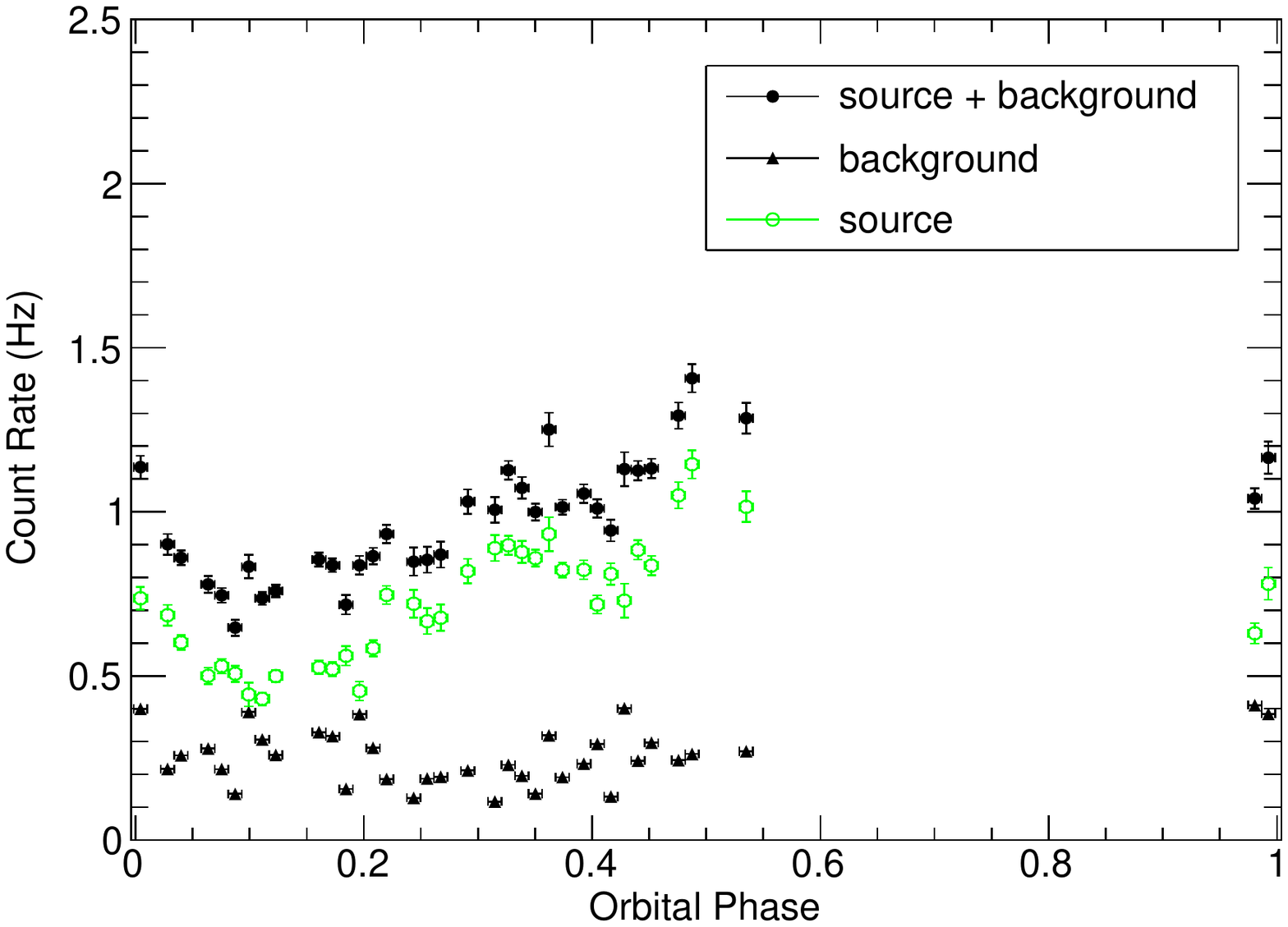}{0.4\textwidth}{(b) OBSID 2030170102--4}}
\gridline{\fig{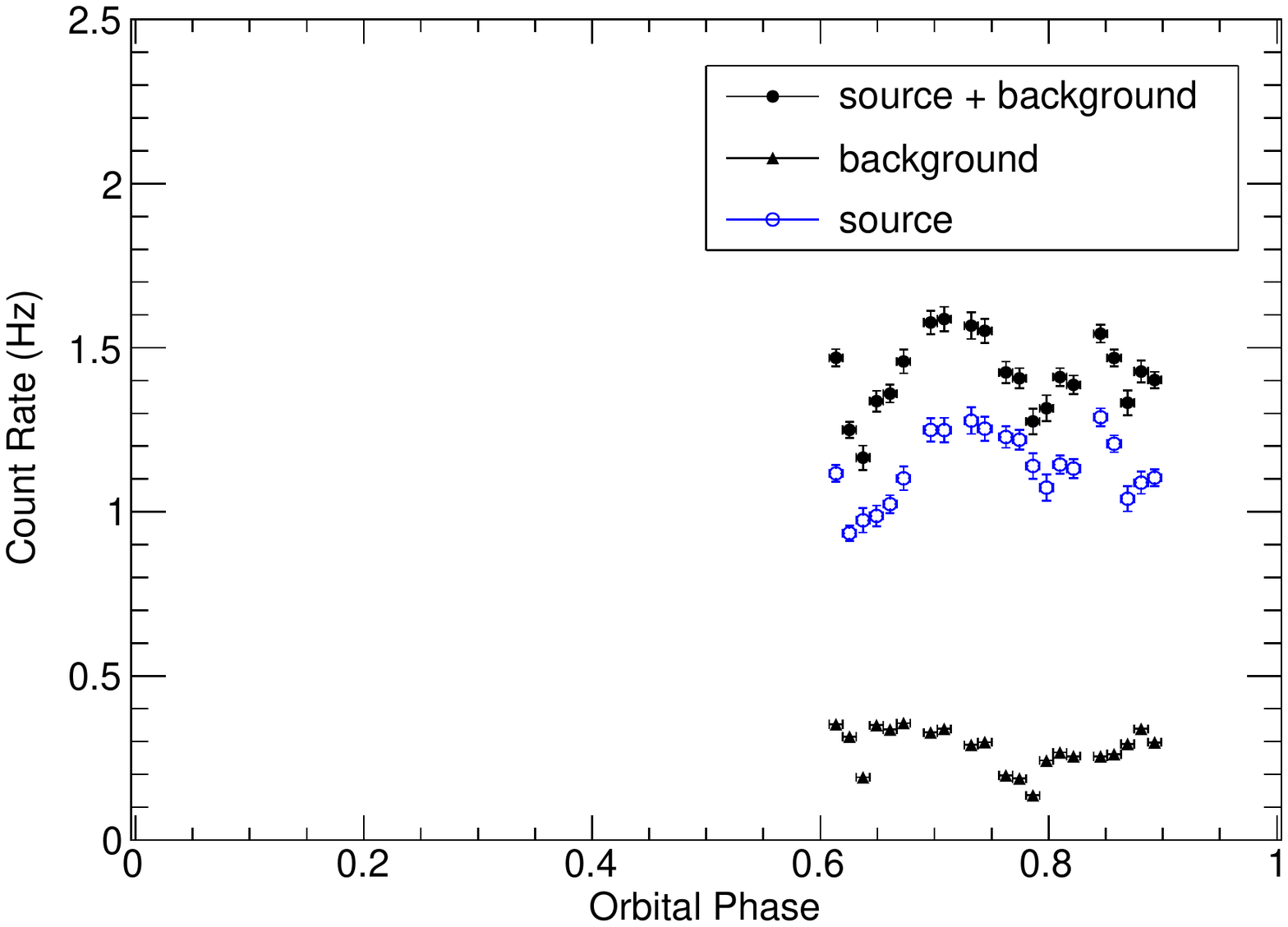}{0.4\textwidth}{(c) OBSID 4631010101--2}
          \fig{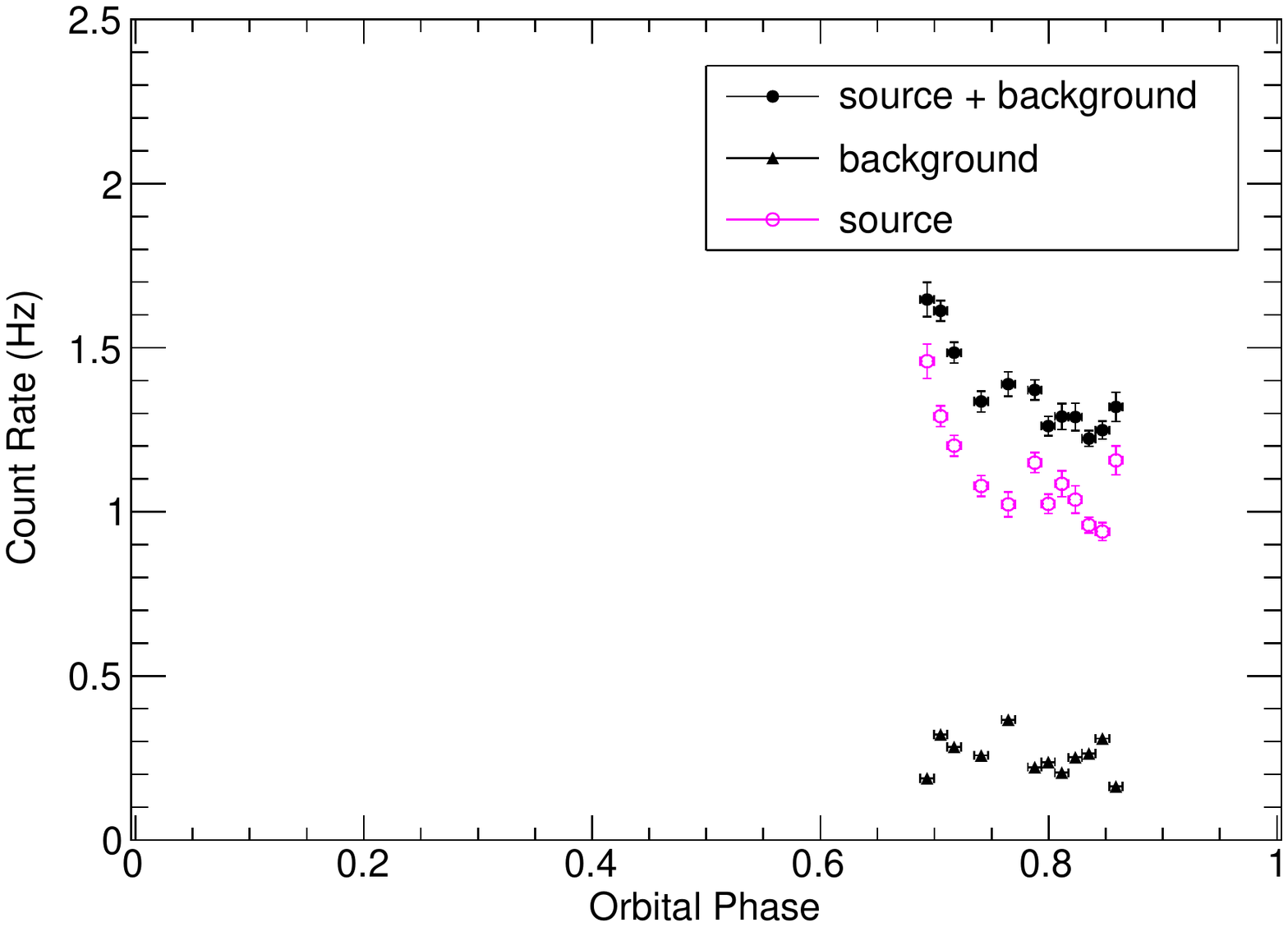}{0.4\textwidth}{(d) OBSID 4631010201--2}}
\caption{The source and background count rates for each \nic observation.}
\label{fig_background_rate}
\end{figure*}

In order to check whether the background events do not cause the reported short-term variability accidentally, we show the light curve of the count rate for both source and background events in Figure~\ref{fig_background_rate}. While the background rate sometimes is comparable to the source rate, the background light curve does not show a feature similar to the reported variability with a time scale of a few tens of ks. Thus, we conclude that the observed short-term variability is not an artifact by the background events.

\end{document}